\documentclass[a4paper,11pt]{article}
\pdfoutput=1 

\usepackage{jheppub1} 

\usepackage[T1]{fontenc} 
\usepackage{amsmath,amssymb,amsfonts,graphicx,subfigure,bm}

\usepackage{natbib, color}
\usepackage{amssymb}
\usepackage{subfigure}
\usepackage{amsfonts}
\usepackage{multicol}
\usepackage{mathtools}
\usepackage{physics}
\usepackage{amsmath}
\usepackage{mathrsfs}
\usepackage{enumitem}
\usepackage{adjustbox}
\usepackage{tensor}
\usepackage[utf8]{inputenc}
\usepackage{titlesec}
\usepackage{indentfirst}
\usepackage{graphicx}

\title{Topological entanglement entropy for torus-knot bipartitions and the Verlinde-like formulas}

\author[]{Chih-Yu Lo, }
\author[]{Po-Yao Chang}

\affiliation[]{Department of Physics, National Tsing Hua University, Hsinchu 30013, Taiwan} 

\emailAdd{chihyulo.jared@gmail.com, pychang@phys.nthu.edu.tw}

\date{}

\numberwithin{equation}{section}

\abstract
{The topological R\'enyi and entanglement entropies depend on the bipartition of the manifold and the choice of the ground states. However, these entanglement quantities remain invariant under a coordinate transformation when the bipartition also undergoes the identical transformation. In topological quantum field theories (TQFTs), these coordinate transformations reduce to representations of the mapping class group on the manifold of the Hilbert space. We employ this invariant property of the R\'enyi and entanglement entropies under coordinate transformations for TQFTs in (2 + 1) dimensions on a torus with various bipartitions. By utilizing the replica trick and the surgery method to compute the topological R\'enyi and entanglement entropies, the invariant property results in Verlinde-like formulas. Furthermore, for the bipartition with interfaces as two non-intersecting torus knots, an $SL(2, \mathbb{Z})$ transformation can untwist the torus knots, leading to a simple bipartition with an effective ground state. This invariant property allows us to demonstrate that the topological entanglement entropy has a lower bound $-2 \ln D$, where $D$ is the total quantum dimensions of the system. }

\begin{document}
\maketitle
\flushbottom


\tableofcontents

\section{Introduction}
After the seminal works on the Chern-Simon theory~\cite{TQFT1,TQFT2}, topological quantum field theories (TQFTs) have found applications in studying the properties of topologically ordered systems with a mass gap. Notable examples include the fractional quantum Hall effect~\cite{FQH, FQH2, FQH3, FQH4,FQH5,FQH6,FQH7,FQH8,FQH9}, gapped quantum spin liquids~\cite{SpinLiquid,SpinLiquid2}, quantum dimer models~\cite{Dimer}, and $p_x + i p_y$ superconductors~\cite{px+ipy}. TQFTs can be regarded as effective field theories for these systems, treating the mass gap as infinite. Under this stringent condition, all excitations become irrelevant, allowing us to focus solely on the ground states, which exhibit long-range entanglement. Due to these long-range entangled properties, the number of ground states depends on the topology of the manifold, leading to these systems being referred to as topological orders~\cite{TO}.

The concept of topological entanglement entropy (TEE) was later introduced in Refs.~\cite{TEE,TEE2} as a diagnostic tool for discerning topological orders in two dimensions. The entanglement entropy of these systems can be written as $\alpha L + S_{\text{TEE}}$, where $\alpha L$ is proportional to the area of the entanglement interface~\cite{Area}, and the $S_{\text{TEE}}$ is a negative term referred to as the TEE due to the topological constraints of these systems. Extensive computations of the TEEs in various models including the fractional quantum Hall effect~\cite{FQHTO1,FQHTO2,FQHTO3,FQHTO4}, gapped quantum spin liquids~\cite{SpinLiquidTO1,SpinLiquidTO2}, toric codes~\cite{ToricCodeTO}, and quantum dimer models~\cite{DimmerTO} have substantiated its efficacy in characterizing topological orders. On the other hand, the TEEs can be directly computed using the Von-Neumann entropy within the framework of TQFTs through the replica method~\cite{TEEinCS}, which successfully reproduces the negative TEEs.
Furthermore, by employing the replica method in TQFTs, several entanglement quantities including the negativity~\cite{negativity, diagTEE, EdgeTEE}, pseudo entropy~\cite{psuedoTEE}, mutual information~\cite{negativity, EdgeTEE} and the reflective entropy~\cite{diagTEE} are investigated.
Moreover, the TEEs can be derived using the diagrammatic approach, which captures the TEEs generated by the braiding of anyons~\cite{anyonTEE, diagTEE, multiTEE}. Additionally, the area term of the entanglement entropy in the context of TQFTs can be recovered by considering a regularization of the entanglement interface~\cite{EdgeTEE, interfaceTEE, diagTEE}.

For a spatial bipartition where subsystem $A$ forms an annulus on a torus in $(2+1)$ dimensions, the TEE associated with such bipartition is proposed \cite{TEE,TEE2} to be given by $-2\ln D$. This matches the calculation  of the TEE for the bipartition with two contractible $S^1$ entanglement interfaces~\cite{TEEinCS,EdgeTEE}.
However, for the two $S^1$ entanglement interfaces being non-contractible loops of the torus, the TEE can be different for different ground states depending on  the absence or presence of the Wilson lines in the bulk~\cite{TEEinCS,gsTEE,EdgeTEE}.
In this paper, we demonstrate that if one fixes a ground state, the TEE can still differ depending on the twists of the annulus. We examine the TEEs for different bipartitions characterized by two non-intersecting torus-knots interfaces between subsystem $A$ and its complement. We will refer to these bipartitions as torus-knot bipartitions.
We demonstrate that the TEE for a torus-knot bipartition is given by 
\begin{equation}\label{decompTEE}
    S_{\mathrm{TEE}} = -2\ln D + S_{\mathrm{gs}},
\end{equation}
where $S_{\mathrm{gs}}$ is defined in Eq.~(\ref{Eq:gs}),
and $0 \leq S_{\mathrm{gs}} \leq 2\ln D$ is a non-negative value that doesn't change the non-positive nature of total TEE. This is similar to the TEE for an untwisted annulus bipartition with Wilson lines inserted~\cite{TEEinCS,gsTEE,EdgeTEE}. That is, the twists of bipartition will induce an effective ground state to the system. 

To compute the TEEs for general torus-knot bipartitions, we leverage coordinate transformations on the torus, specifically the mapping class group $SL(2, \mathbb{Z})$ in TQFTs. When the bipartition of the manifold transform with respect to the coordinate transformation, the TEE and other entanglement quantities remain invariant under this transformation. Exploiting this invariant property of entropies, we map torus-knot bipartitions to a simpler annulus bipartition, whose TEE can be easily computed.
As a byproduct of our approach, for the torus-knot bipartition with two $S^1$ interfaces forming the meridians of the torus, the invariance of the R\'enyi entropies under modular $S$ transformation yields Verlinde-like formulas. These formulas establish relations between quantum dimensions and modular data.

This paper is organized as follows. In Sec.~\ref{section:coordinate}, we provide the definition of coordinate transformations, interpreted as relabelings of sites on the manifold. This expression establishes the invariant property of entanglement quantities. Sec.~\ref{Sec:3} reviews the replica trick and the surgery method for computing R\'enyi entropies in TQFTs.
In Sec.~\ref{Sec:4}, we apply the invariant property of R\'enyi entropies under coordinate transformations with various annulus bipartitions, including those with meridian interfaces, longitude interfaces, and general torus knot interfaces. The invariance of R\'enyi entropies is demonstrated to lead to Verlinde-like formulas. Additionally, we demonstrate that torus-knot bipartitions give rise to an effective ground state.
Finally, we summarize our findings and conclude the paper in Section~\ref{Sec:5}.

\section{Coordinate transformation}\label{section:coordinate}

\begin{figure}
 \centering
\includegraphics[height = 6cm]{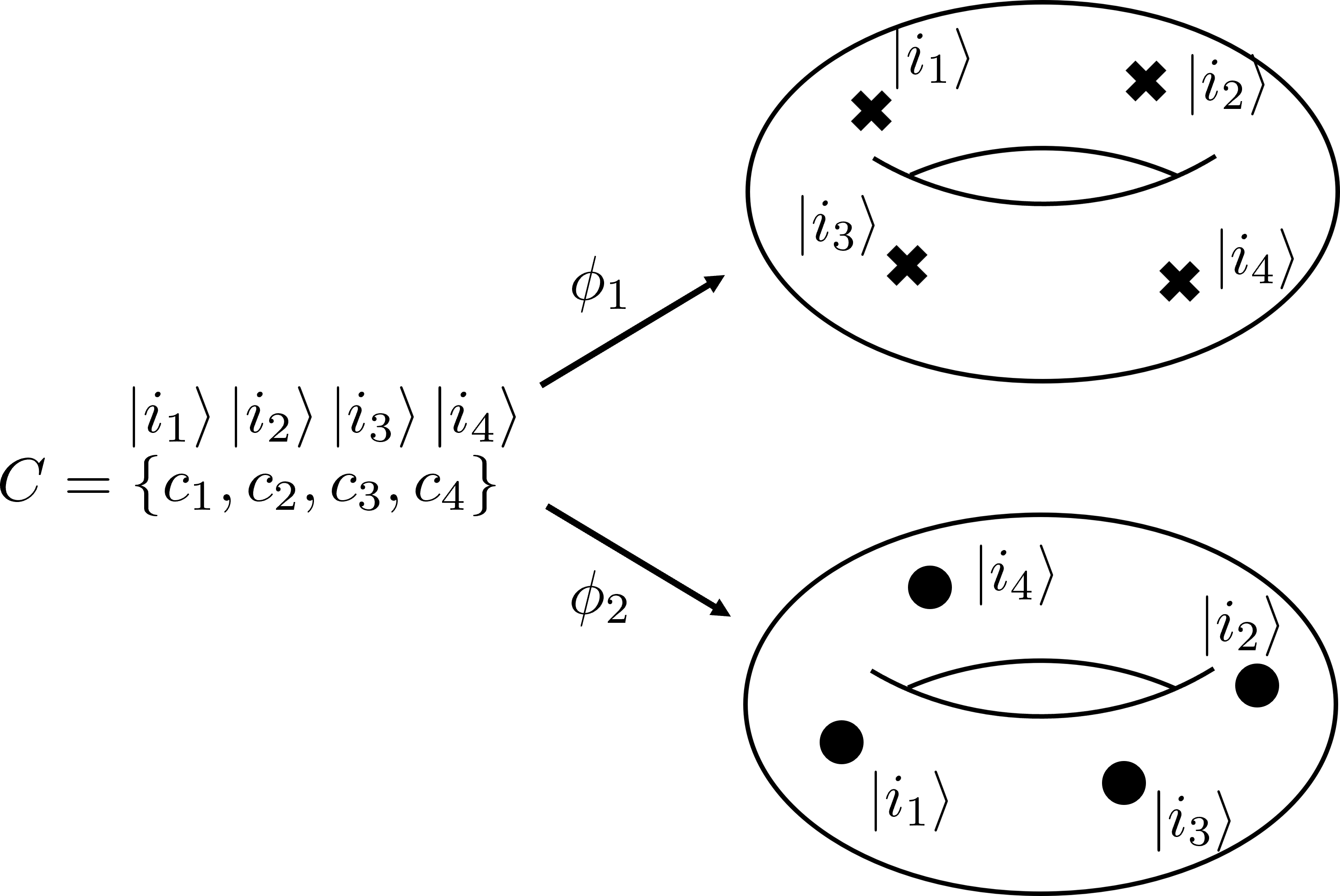}    
\caption{Two different maps $\phi_1$ and $\phi_2$ from sites to space resembling the same state.}
\label{F1}
\end{figure}

Before delving into the calculation of the TEEs, let us discuss the invariance property under coordinate transformations. Suppose we have a system of sites indexed by a set $C$ placed on a surface $S$ through a bijective morphism\footnote{For example, it preserves topology for topological manifolds or smoothness for smooth manifolds.} 
$\phi: C \rightarrow S$. The Hilbert space of the system is given by
\begin{equation*}
    \mathcal{H}^{C} =  \otimes_{c \in C} \mathcal{H}^c_{\phi(c)} = \otimes_{x \in S} \mathcal{H}^{\phi^{-1}(x)}_x = \mathcal{H}_S.
\end{equation*}
In the context of quantum field theories, $\mathcal{H}^c_{\phi(c)}$ represents the same Hilbert space, independent of site indices. Let ${\ket{i}^c_{\phi(c)} }{i \in J}$ be an orthonormal basis for $\mathcal{H}^c_{\phi(c)}$ (possibly infinite-dimensional in the case of bosons). Then, a basis for $\mathcal{H}_S$ can be expressed as
\begin{equation}\label{def}
    \{ \ket{f}_{\phi}^C = \otimes_c \ket{f(c)}^{c}_{\phi(c)} = \otimes_x \ket{f\circ \phi^{-1}(x)}^{\phi^{-1}(x)}_{x} | f \in F(C,J)  \},
\end{equation}
where $F(C,J)$ denotes the set of functions from $C$ to $J$. The label $\phi$ indicates the position of sites, and the label $C$ indicates that it is the functional basis ranged in $C$. We define our states as intrinsic objects attached to sites, independent of spatial labelings [See Fig.~\ref{F1}]. That is, for any given labelings $\phi_1, \phi_2 : C \rightarrow S$,
\begin{equation}\label{citeeq}
    \ket{i}^c_{\phi_1(c)} = \ket{i}^c_{\phi_2(c)} \quad  \text{and}\,\, \quad \ket{f}^C_{\phi_1} = \ket{f}^C_{\phi_2}.
\end{equation}

Now, a state in $\mathcal{H}^C$ can be expressed as
\begin{equation*}
    \ket{\psi^C}_{\phi} = \sum_{f\in F(C,J)} \psi^C(f) \ket{f}^C_{\phi},
\end{equation*}
where the functionals $\psi^C : F(C,J) \rightarrow \mathbb{C}$ satisfy the equation of motion for the underlying system.

Let us consider a bipartition on sites $C = A \cup B$ and let $\{\ket{f^A}^C_{\phi}\}_{f^A \in F(A,I)}$, $\{ \ket{f^B}^C_{\phi}\}_{f^B \in F(B,I)}$ be the bases for $\mathcal{H}^A, \mathcal{H}^B$ respectively. A continuous function $f$ can be bijectively mapped to continuous functions $f^A, f^B$ on $A$ and $B$ by restriction.
Therefore, we write $\ket{f}^C_{\phi} = \ket{f^A}^C_{\phi} \otimes \ket{f^B}^C_{\phi}$. 
With this identification, we define $\psi^{AB}: F(A,J) \times F(B,J) \rightarrow \mathbb{C}$ by 
\begin{equation*}
    \psi^{AB}(f^A, f^B) = \psi^C(f).
\end{equation*}
Now, we have 
\begin{equation}
    \ket{\psi^C}_{\phi} = \sum_{f} \psi^C(f) \ket{f}^C_{\phi}= \sum_{f^A,f^B} \psi^{AB}(f_A,f_B) \ket{f^A}^A_{\phi} \otimes \ket{f^B}^B_{\phi},
\end{equation}
and
\begin{equation}
    \rho_A = \sum_{f^A, f^{A'}} \Big(\sum_{f^B} \psi^{AB}(f_A,f_B) \Bar{\psi}^{A'B}(f_{A'},f_B) \Big) \ket{f^A}^A \langle f^{A'}|^{A'} .
\end{equation}

Suppose we perform a coordinate transformation on $S$, which is an automorphism $K:S \rightarrow S$ on the manifold\footnote{For example, we can change the Cartesian coordinate to the polar coordinate. A more exotic example is a Dehn twist on a torus, where the coordinate transformation can be considered as the twisting map $[e^{i \theta}, t] \to [e^{i( \theta+2\pi t)},t]$ of the annulus with $\theta \in [0 ,2\pi]$ and $t\in[0,1]$.}. The Hilbert space can now be written as
\begin{equation*}
    \mathcal{H}^C =  \otimes_{c \in C} \mathcal{H}^c_{K\circ \phi(c)} = \otimes_{x \in S} \mathcal{H}^{(K\circ \phi)^{-1}(x)}_x = \mathcal{H}_S.
\end{equation*}
Under such an automorphism, everything discussed above remains the same with the substitution $\phi \rightarrow K \circ \phi$. Essentially, we are changing nothing but shuffling the position of sites around. Therefore, if we perform a partial trace with respect to the relabeling of the site labels, the reduced density matrix will be identical.

\begin{figure}
\centering
\includegraphics[height = 3.5cm]{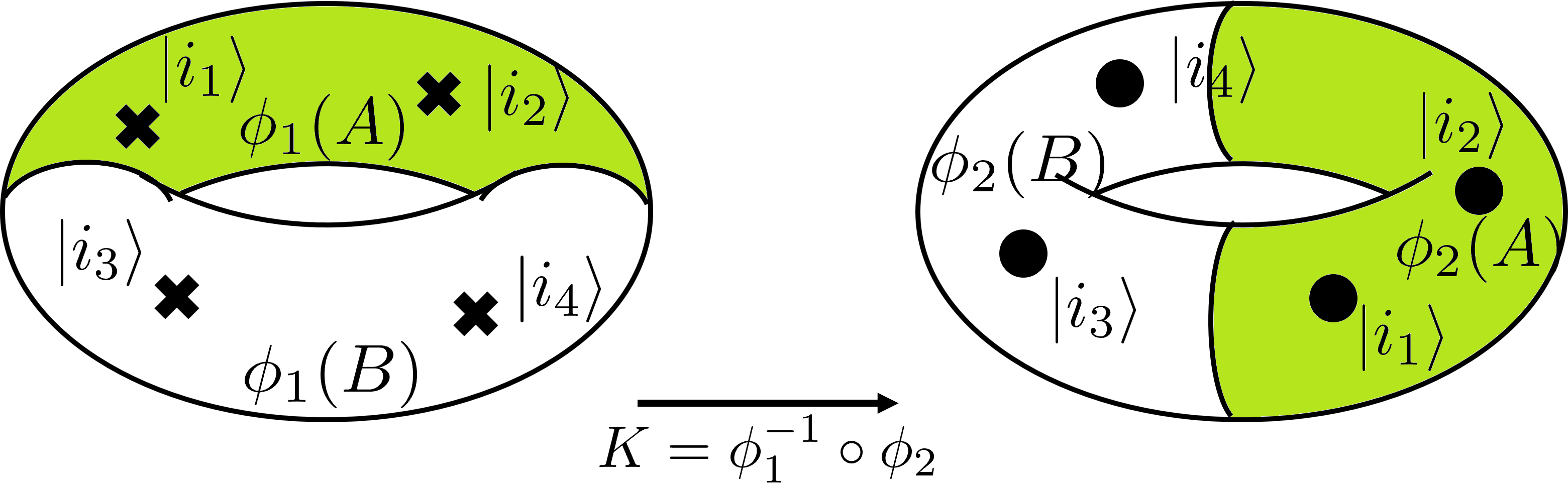}
\caption{Same bipartition on sites can gives different bipartitions on space}
\label{F2}
\end{figure}

On the other hand, we can also look at the state in the position $x$. By the definition Eq. (\ref{def}), we have the relation
\begin{equation}\label{cite-position}
    \ket{f}^C_\phi = \ket{f\circ \phi^{-1}}^\phi_S.
\end{equation}
One should note that there can be ambiguity in defining states in positions in quantum field theories\footnote{There is no unique way to discretize continuous theories, and the states can depend on the regularization scheme.}, and in general, $\ket{i}_x^{\phi^{-1}_1(x)} \neq \ket{i}^{\phi^{-1}_2(x)}_x$ and $\ket{f}^{\phi_1}_S \neq \ket{f}^{\phi_2}_S$. In fact, we can define an automorphism on $S$ by $\phi_2 \circ \phi_1^{-1}$, then
\begin{equation}
    \ket{f}^{\phi_2}_S = \sum_{f'} U_{\phi_2 \circ \phi_1^{-1}}(f,f') \ket{f'}^{\phi_1}_S,
\end{equation}
where $U$ is an unitary operator on the functional space.
This can be think of as performing a relabeling on sites while keeping the spatial wave functional basis fixed.
On the other hand, from Eq.~(\ref{citeeq}) and Eq.~(\ref{cite-position}) we have $\ket{f}^{\phi}_S = \ket{f\circ K^{-1}}^{K \circ \phi}_S$, thus
\begin{equation}
    \ket{f\circ K}^{\phi}_S =  \sum_{f'} U_{K}(f,f') \ket{f'}^{\phi}_S.
\end{equation}
This is essentially changing the spatial functional while keeping the site labels fixed. A change of labeling of sites can be thought of as a two-step process. The unitary transformation cancels out on each other, leaving the state invariant. More explicitly, define $K = \phi_1 \circ \phi^{-1}_2$, then
\begin{equation}
    \ket{f}^C_{\phi_1} = \ket{f}^C_{K \circ \phi_2} = \ket{f\circ K^{-1}}^{\phi_1}_S = \sum_{f', f''}  U_{K^{-1}}(f,f')U_{K}(f',f'') \ket{f''}^{\phi_2}_S = \ket{f}^C_{\phi_2} .
\end{equation}
It is crucial to note that while the bipartition with respect to the sites is fixed, the bipartition with respect to the positions will also change according to the relabeling of sites $S = \phi_1(A) \cup \phi_1(B) \rightarrow S = \phi_2(A) \cup \phi_2(B)$ [See Fig.~\ref{F2} for example].

In the case of TQFTs, the Hilbert space of functionals is reduced to a finite-dimensional Hilbert space of anyons if the gauge group and the manifold are compact~\cite{TQFT1, TQFT2}. The coordinate transformations are also replaced by the mapping class group, which consists of topologically equivalent classes of transformations.

\section{Partition functions, replica trick, and the surgery method}
\label{Sec:3}
Here, we briefly introduce the method of computing the TEEs in the context of TQFTs following Ref.~\cite{TEEinCS}. Let us consider the Chern-Simon action on a three manifold $M$
\begin{equation}
    S_{\mathrm{CS}} = \frac{k}{4\pi} \int_M \mathrm{Tr}(A \wedge dA + A \wedge A \wedge A), 
\end{equation}
where $A$ is the connection one-form of the principal bundle of the underlying gauge group, and $k$ is the level of the theory. Doing a path integral over $M$ will give rise to a state in the Hilbert space of its boundary,
\begin{equation}
    \ket{M} = \int \mathcal{D}A \, e^{-i S_{\mathrm{CS}}} \in \mathcal{H}_{\partial M}.
\end{equation}
In general, one can also insert Wilson loops into the bulk to obtain different boundary states
\begin{equation}
    \ket{M,C_a} = \int \mathcal{D}A \, e^{-i S_{\mathrm{CS}}[A]} W_a^C[A],
\end{equation}
where the Wilson loop operator
\begin{equation}
    W_a^C[A] = \mathrm{Tr}_a \mathcal{P}\{e^{\oint_C A} \}
\end{equation}
traces the holonomy on a closed curve $C$ in representation $a$.\par

In particular, $\mathcal{H}_{T^2}$ is spanned by $\ket{a} = \ket{D^2 \times S^1, C_a}$, with $a$ ranging over the highest weight representations of the gauge group at level $k$~\cite{Ver,TQFT1,TQFT2}. One can also construct the dual vectors by reversing the orientation of the manifold
\begin{equation}
    \bra{a} = | \overline{D^2 \times S^1} , \overline{C_a} \rangle \in \mathcal{H}^*_{T^2} =  \mathcal{H}_{\overline{T^2}}.
\end{equation}
Partition function on $S^2 \times S^1$ can be obtained by gluing two $D^2 \times S^1$ with opposite orientation by the identity map
\begin{equation}\label{eqn:Iab}
    Z(S^2 \times S^1, C_{\bar{a}}, C_b) = \bra{a} \ket{b} = \delta_{ab}.
\end{equation}
On the other hand, one can also obtain the partition function of $S^3$ by gluing two $D^2 \times S^1$ with opposite orientation related by the modular $S$ transformation
\begin{equation}\label{eqn:Sab}
    Z(S^3, L_{ab}) = \bra{a} S \ket{b} = S_{ab},
\end{equation}
where $L_{ab}$ is the configuration of $C_{\bar{a}}$ and $C_b$ being linked once \adjustbox{valign = c}{\includegraphics[height = 1.5cm]{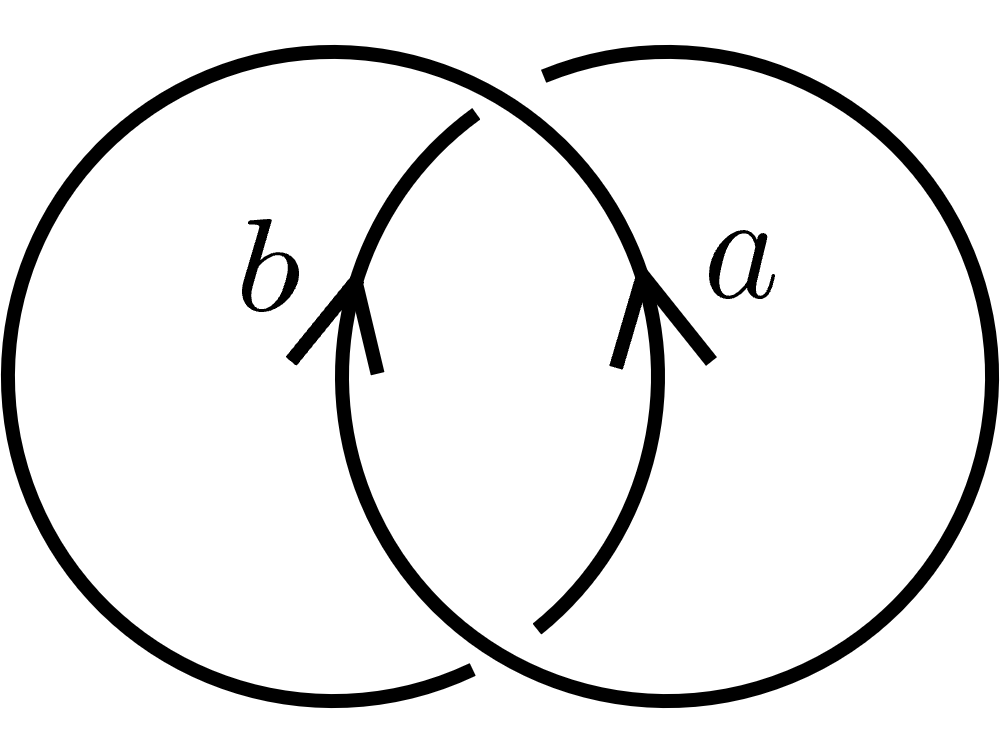}}.
In particular, if $a=b=0$, we obtain $Z(S^2 \times S^1) = 1$ and $Z(S^3) = S_{00} = D^{-1}$. Throughout this paper, we will refer to $\{\ket{a}\}_a$ as the inside basis and $\{S \ket{a}\}_a$ as the outside basis.\footnote{One can imagine that a torus embedded in $\mathbb{R}^3\cup {\infty} \simeq S^3$ divides the ambient space into inside and outside solid tori.} We then have the basis transformation
\begin{equation}\label{basistrans}
    S \ket{a} = \sum_b S_{ab} \ket{b}.
\end{equation}
\par

Given a state $\ket{M}$, one can obtain the corresponding density matrix $\rho$ by the disjoint union of $M$ with $\overline{M}$. Given a bipartition $\partial M = A \cup B$, one can obtain the reduced density matrix $\rho_A = \mathrm{Tr}_B \rho$ by gluing $B$ of $M$ to the corresponding $B'$ of $\overline{M}$ by the identity map. $\rho_A$ is a manifold with boundary $A \cup A'$, where $A'$ is the corresponding region to $A$ in $\overline{M}$. 
One can then obtain $\mathrm{Tr} \rho_A^2$ by making two copies of $\rho_A$ and then perform two gluings from $A$ to $A'$, which is a manifold without boundary. 
Similarly, $\mathrm{Tr} \rho_A^n$ can be constructed by $n$ copies of $M$ with corresponding gluings of the boundaries. The R\'enyi entropies are
\begin{equation}
    S_n = \frac{1}{1-n} \ln \mathrm{Tr} (\rho_A)^n,
\end{equation}
and the entanglement entropy is the limit $n\rightarrow 1$ by performing analytical continuation
\begin{equation}
    S_{\mathrm{TEE}} = \lim_{n\rightarrow 1} S_n.
\end{equation}\par
To evaluate the partition function of the glued manifold, one can apply the method of surgery~\cite{TQFT1} to decompose the manifold into simple ones whose partition functions are known as examples in Eq.~(\ref{eqn:Iab}) and Eq.~(\ref{eqn:Sab}).\par
Suppose one has a manifold $M$ without a boundary, which can be obtained by gluing $\overline{M_1}, M_2$ along their common $S^2$ boundary
\begin{equation}
    Z(M) = \bra{M_1}\ket{M_2}.
\end{equation}
Due to the one dimensional property of $\mathcal{H}_{S^2}$, one has
\begin{equation}
    \adjustbox{valign = c}{\includegraphics[height = 4.5cm]{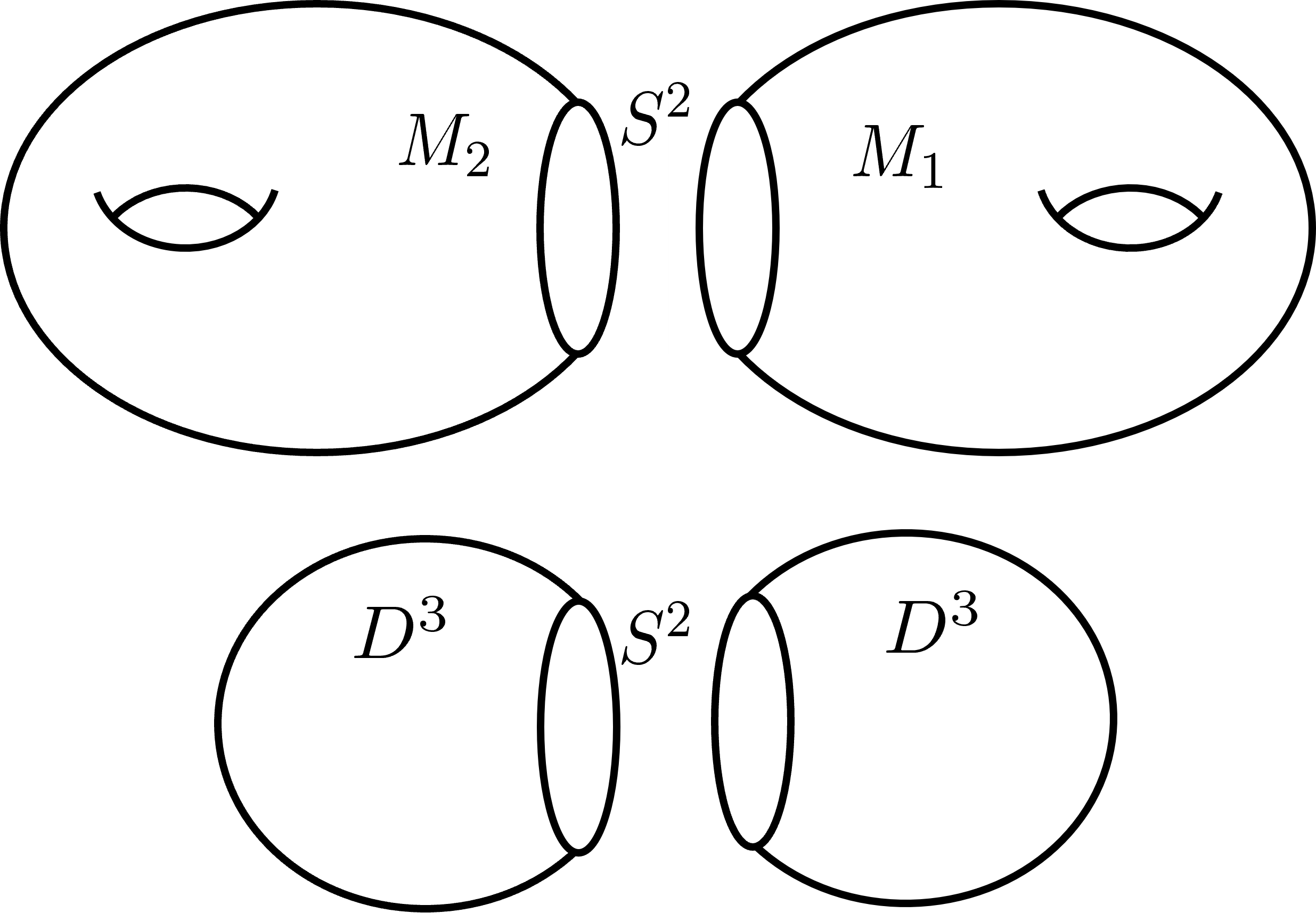}}\hspace{0.1cm}
    = \hspace{0.3cm}
    \adjustbox{valign = c}{\includegraphics[height = 4.5cm]{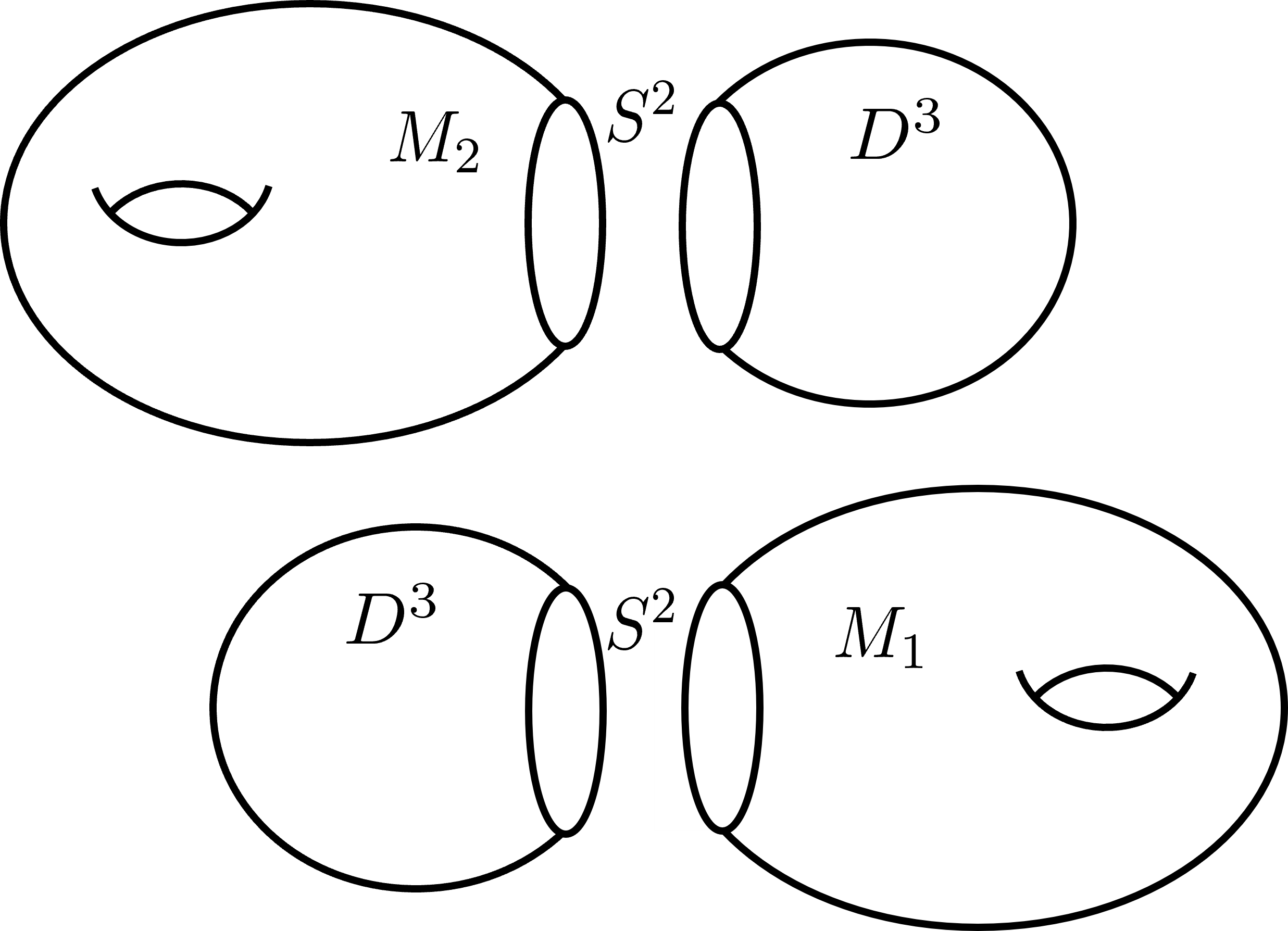}}\hspace{0.1cm}.
\end{equation}
Let $\Tilde{M_1}, \Tilde{M_2}$ be the manifolds obtained by capping off the $S^2$ boundary with $D^3$ of $M_1, M_2$ respectively, we have 
\begin{equation}
    Z(M) Z(S^3) = Z(\Tilde{M_1}) Z(\Tilde{M_2}).
\end{equation}
For example
\begin{equation}\label{eqn:surgery}
    \adjustbox{valign = c}{\includegraphics[height = 2.5cm]{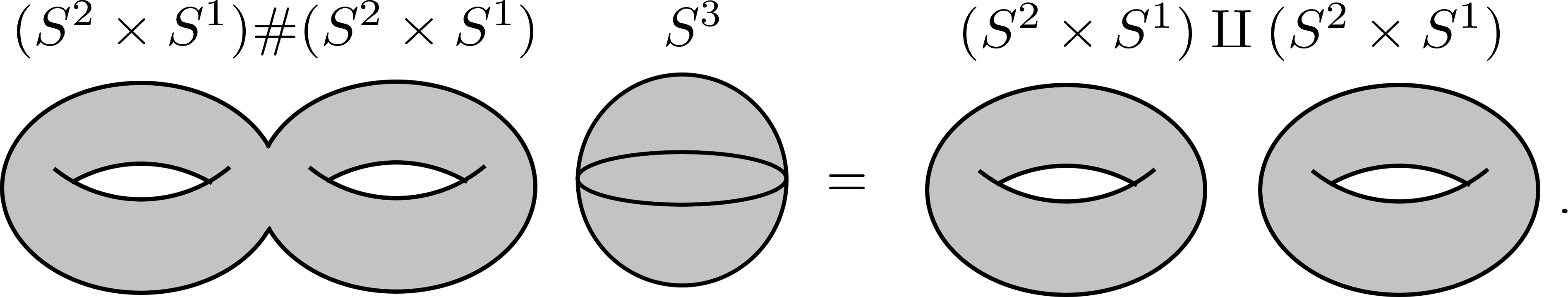}}\hspace{0.1cm}
\end{equation}
In general, one can also perform surgery with a Wilson loop insert due to the one dimensional property of $\mathcal{H}_{S^2, a,\Bar{a}}$ \cite{TQFT1,negativity}
\begin{equation}\label{eqn:surgery_line}
    \adjustbox{valign = c}{\includegraphics[height = 5cm]{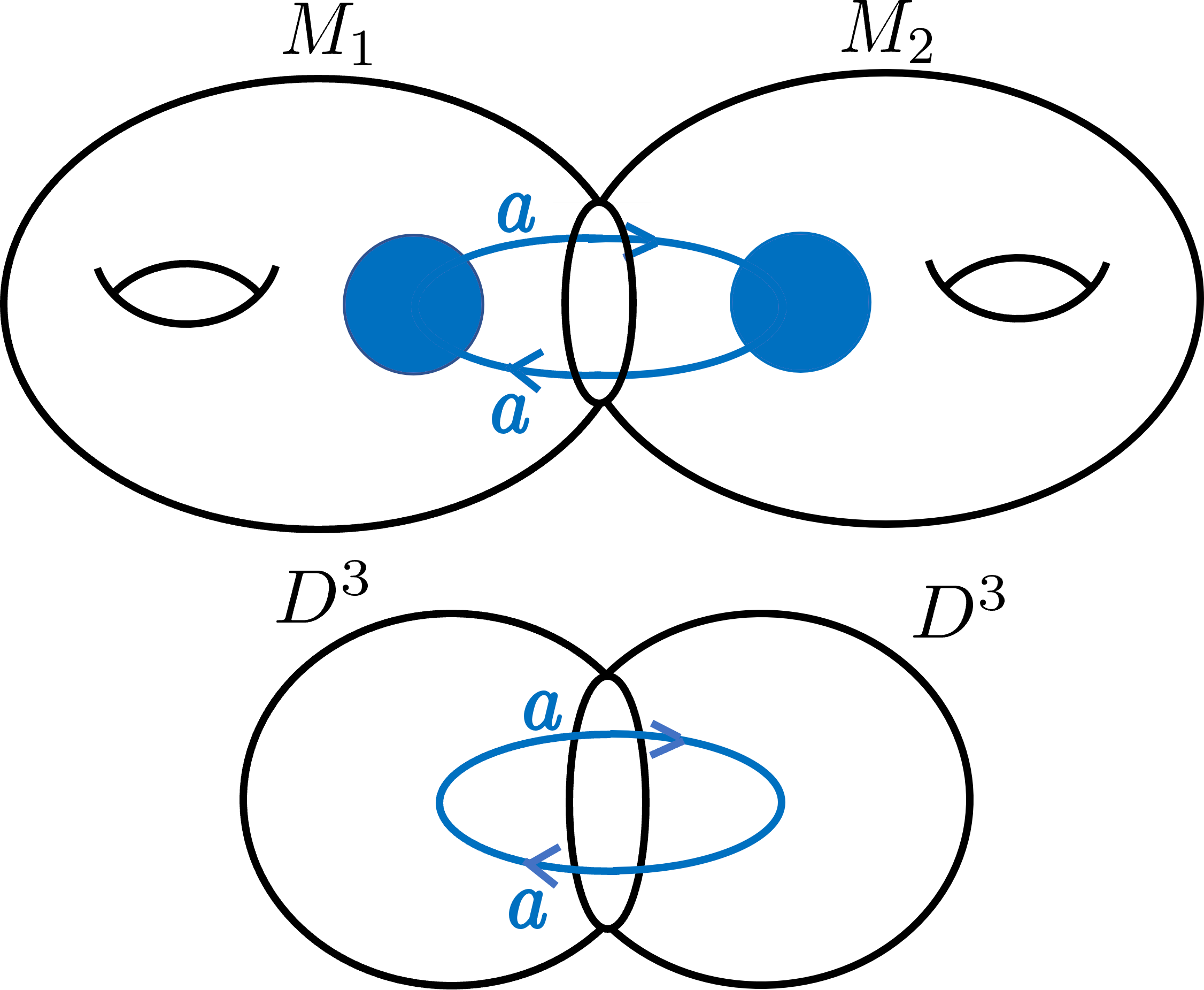}}\hspace{0.1cm}
    = \hspace{0.3cm}
    \adjustbox{valign = c}{\includegraphics[height = 5cm]{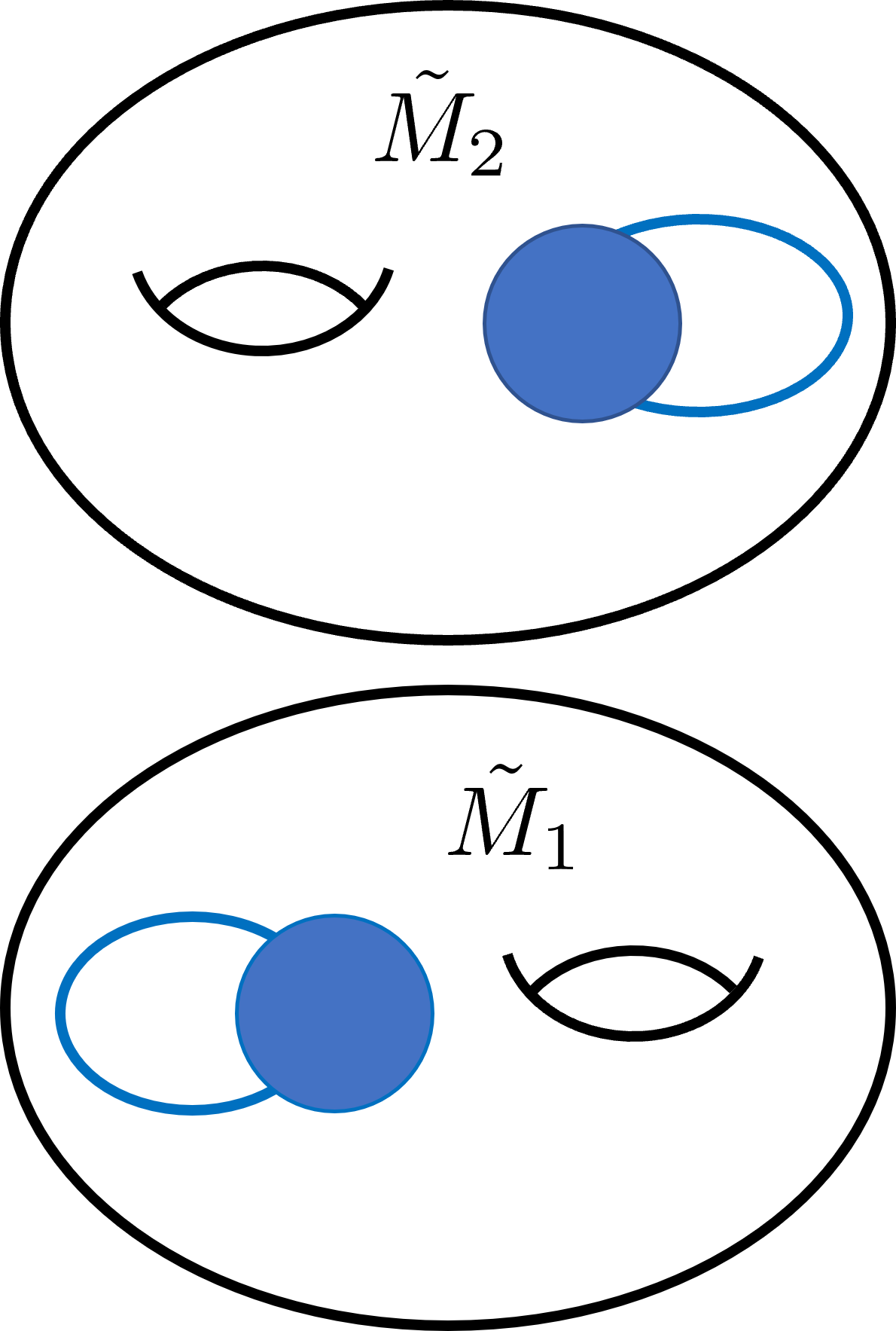}}\hspace{0.1cm},
\end{equation}
 where the blue disks represent arbitrary Wilson line configurations, each potentially different from the others.

\section{Topological entanglement entropy with different bipartitions, and the applications of the coordinate transformation}
\label{Sec:4}

In this section, we apply the invariant property discussed in Sec. \ref{section:coordinate} to compute entropies using different coordinates. In Sec. \ref{subsection4.1}, we will discuss bipartitions with a single $S^1$ interface as a consistency check of the method. In Sec. \ref{subsection4.2}, we will discuss bipartitions with two $S^1$ interfaces being the meridians or the longitudes and obtain Verlinde-like formulas. In Sec. \ref{subsection4.3}, we will discuss the case where the two $S^1$ interfaces are generic torus knots.

\subsection{Single \texorpdfstring{$S^1$}{} interface}\label{subsection4.1}
Consider the vacuum state $\ket{0}$ on $T^2$ generated by the empty solid doughnut $D^2 \times S^1$ with a bipartition which has a single $S^1$ interface 
\begin{equation}
    \ket{0} = \hspace{0.2cm}
    \adjustbox{valign = c}{\includegraphics[height = 2.4cm]{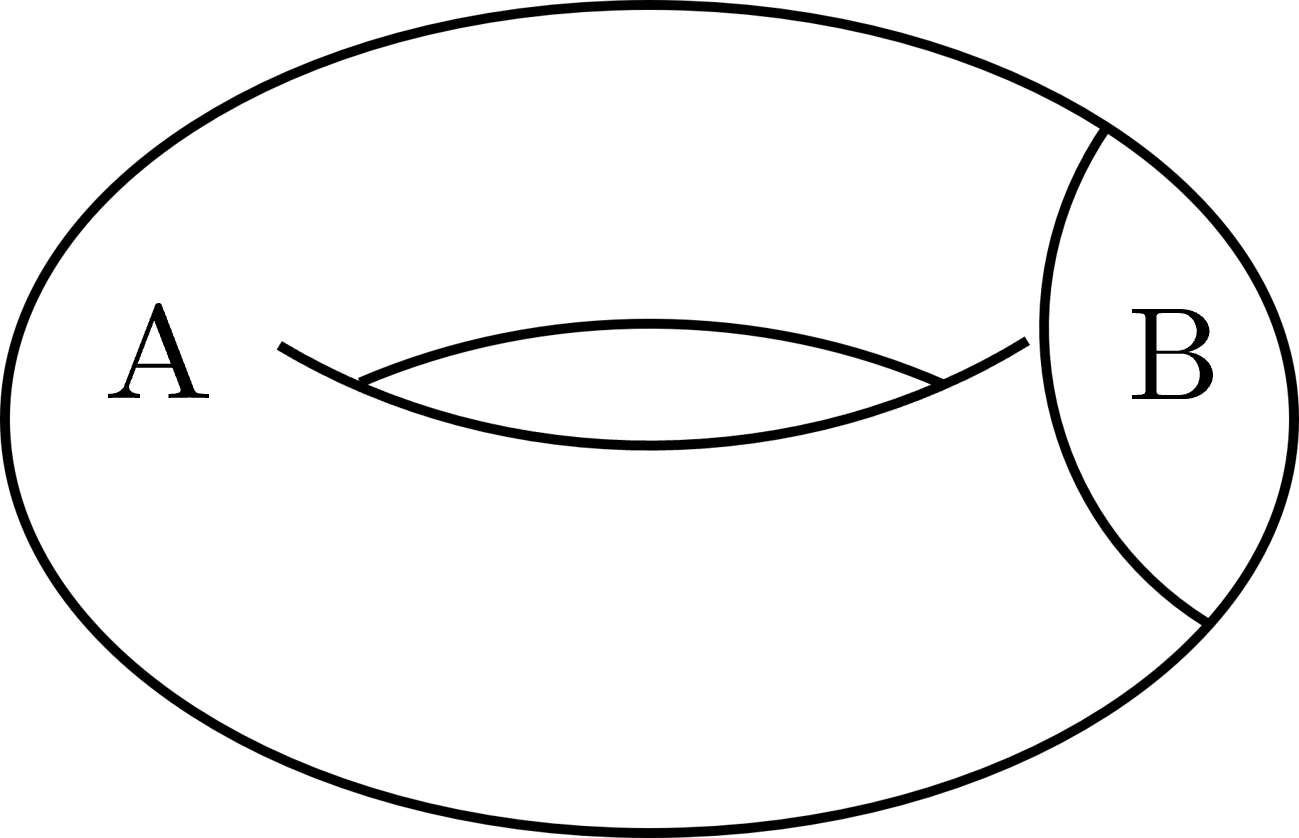}}\hspace{0.1cm}.
\end{equation}
We consider the reduced density matrix
\begin{equation}
    \rho_A = \mathrm{Tr}_B \rho = \mathrm{Tr}_B \ketbra{0}{0} = \hspace{0.2cm}
    \adjustbox{valign = c}{\includegraphics[height = 2.5cm]{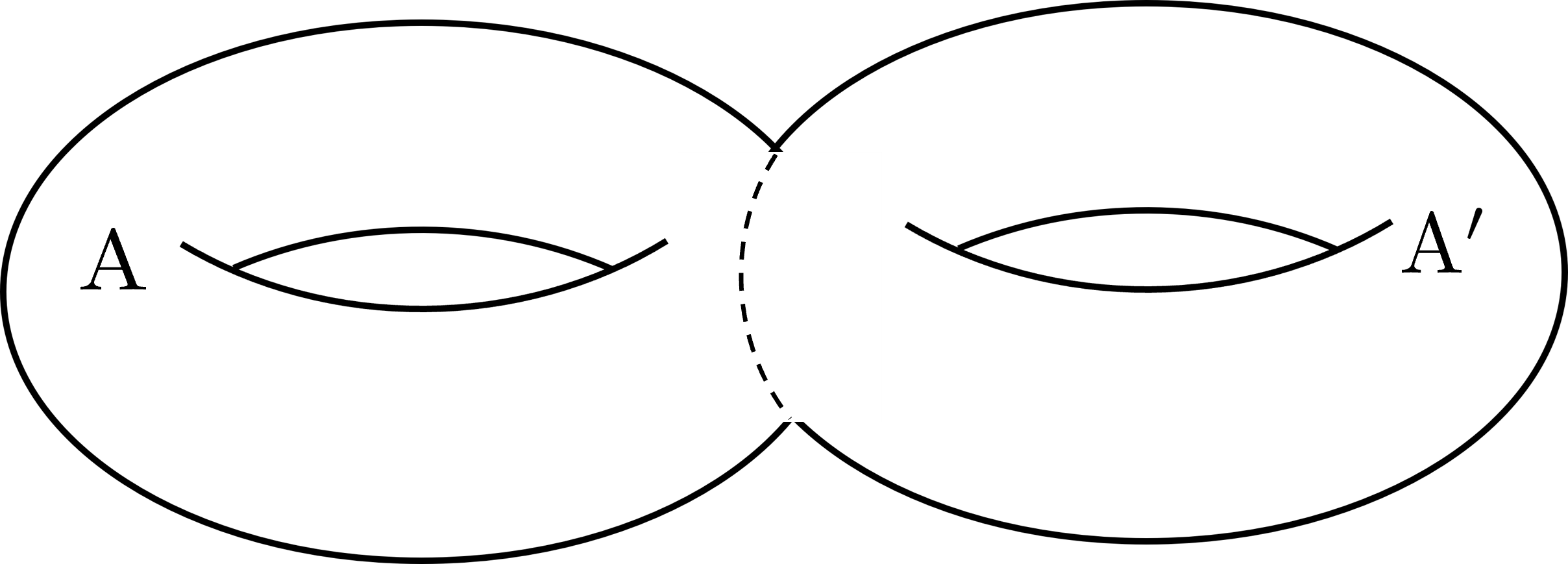}}\hspace{0.1cm}.
\end{equation}
We can compute the $n$-th Rényi entropy by making \(n\) copies of \(\rho_A\) and gluing them together according to the boundary orientation. The \(\mathrm{Tr}{(\rho_A}^n)\) is given by the connected sum of \(n\) copies of \(S^2 \times S^1\).
\begin{equation}
    \mathrm{Tr}{(\rho_A}^n) = \hspace{0.1cm}
    \adjustbox{valign = c}{\includegraphics[height = 2cm]{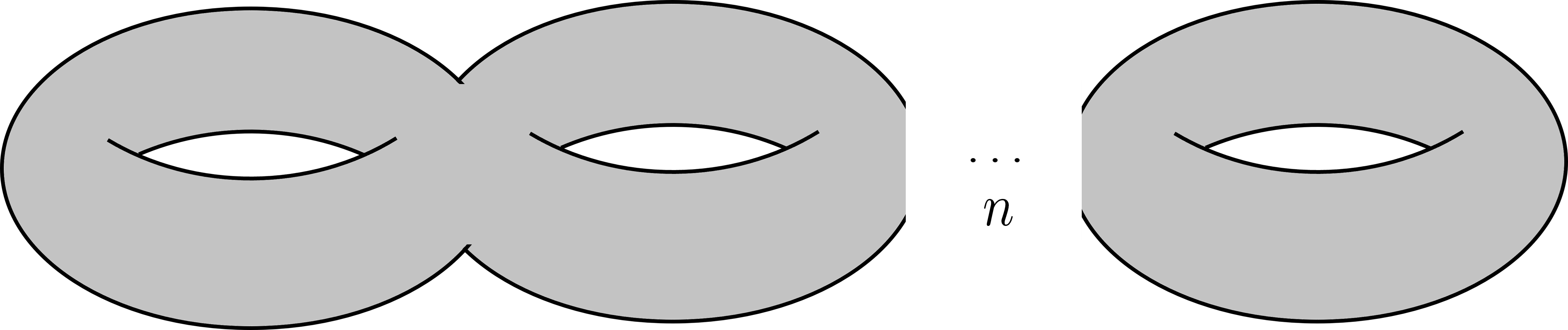}}.
\end{equation}

The partition function of such a configuration can be computed by the method of surgery similar to Eq.~(\ref{eqn:surgery}). By inserting $(n-1)$ copies of $S^3$, the configuration can be separated into $n$ copies of $S^2 \times S^1$. The result is given by
\begin{equation}
\mathrm{Tr}{(\rho_A}^n ) = \frac{Z(S^2 \times S^1)^n}{Z(S^3)^{n-1}} = D^{n-1}.
\end{equation}

On the other hand, we can perform the same calculation using the outside basis. We know from Eq.~(\ref{basistrans}) that the inside torus is related to the outside torus by the transformation
\begin{equation}
    \adjustbox{valign = c}{\includegraphics[height = 2.5cm]{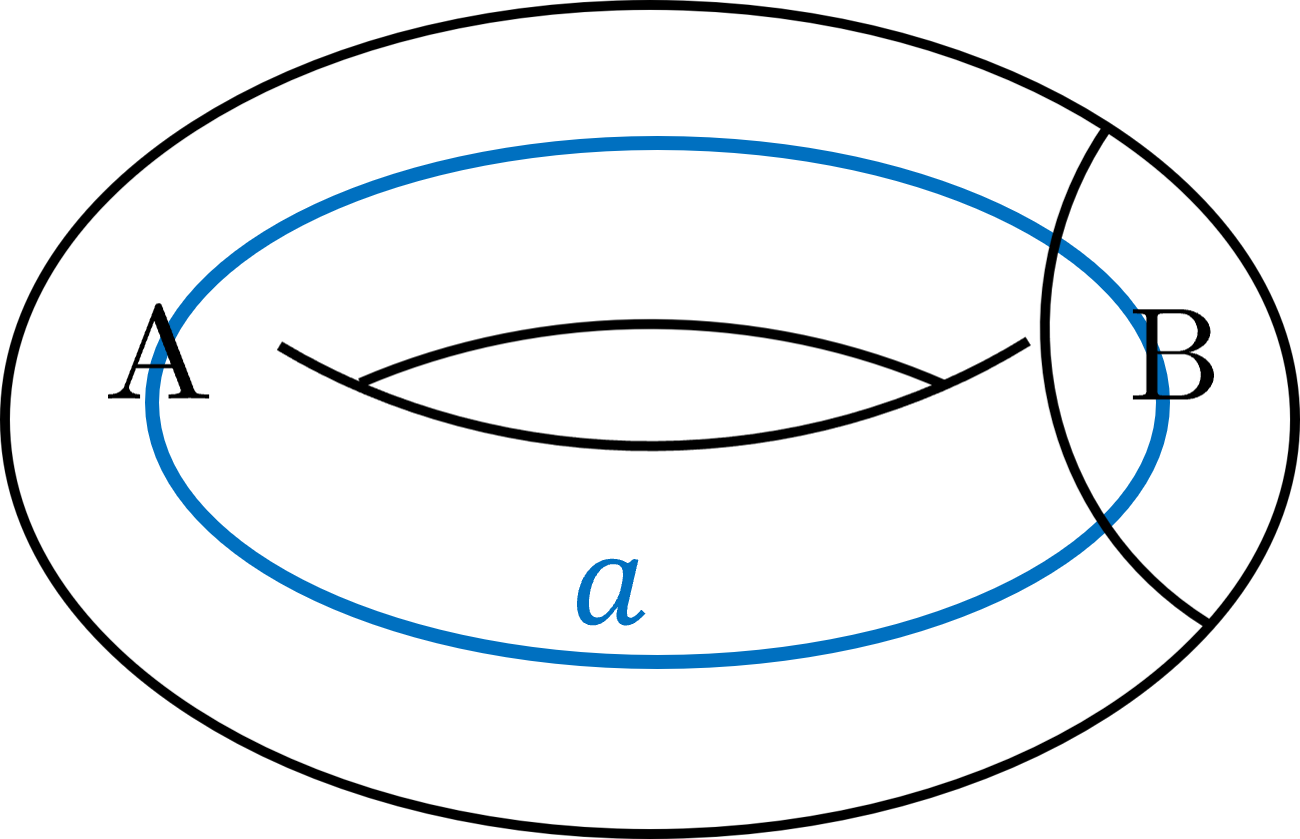}}\hspace{0.1cm}
    = \sum_{b} S_{ab} \hspace{0.2cm}
    \adjustbox{valign = c}{\includegraphics[height = 4cm]{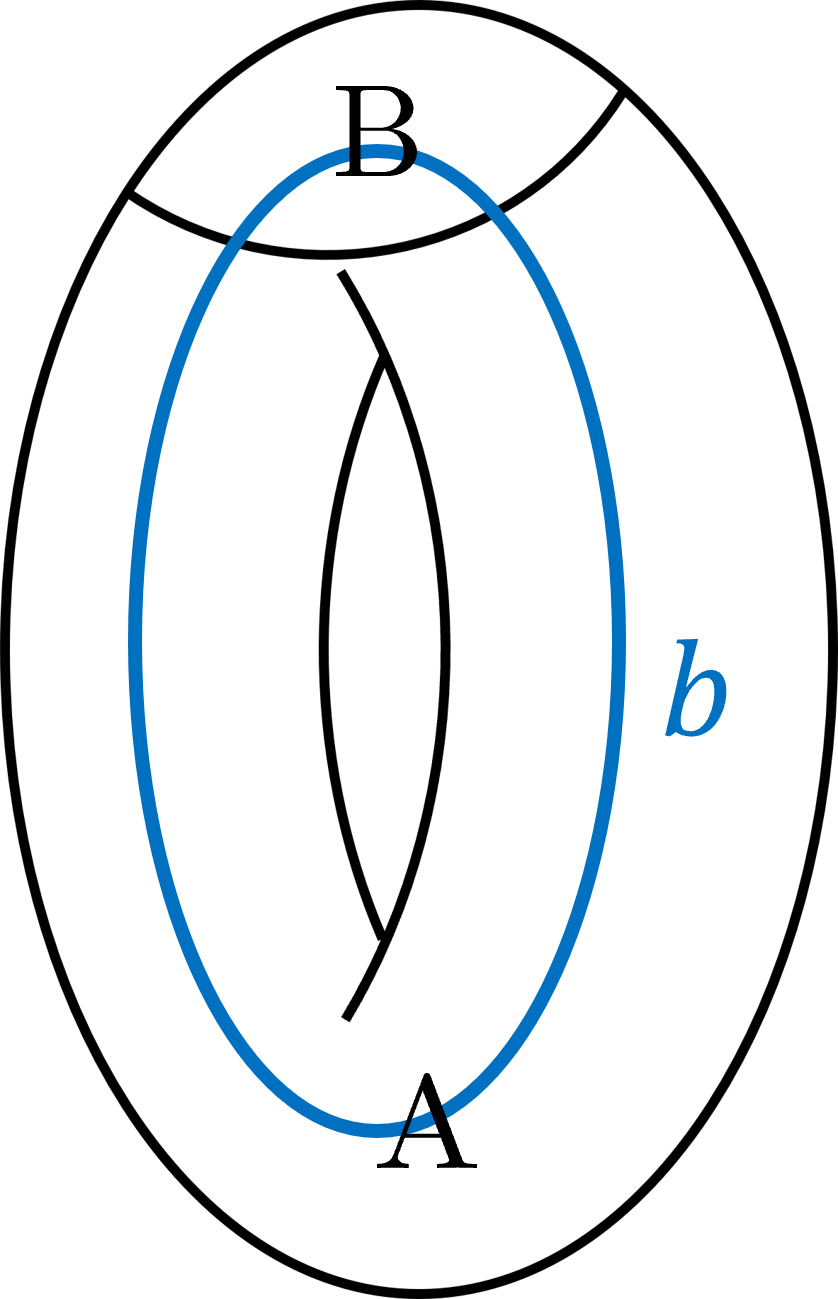}}\hspace{0.1cm},
\end{equation}
where we use vertical torus to indicate the outside basis and the $S_{ab}$ is the modular $S$ matrix.
Using the outside basis, the reduced density matrix is given by
\begin{equation}
    \rho_A = \sum_{a,b} S_{0a} S_{0 \Bar{b}}\hspace{0.2cm}
    \adjustbox{valign = c}{\includegraphics[height = 6cm]{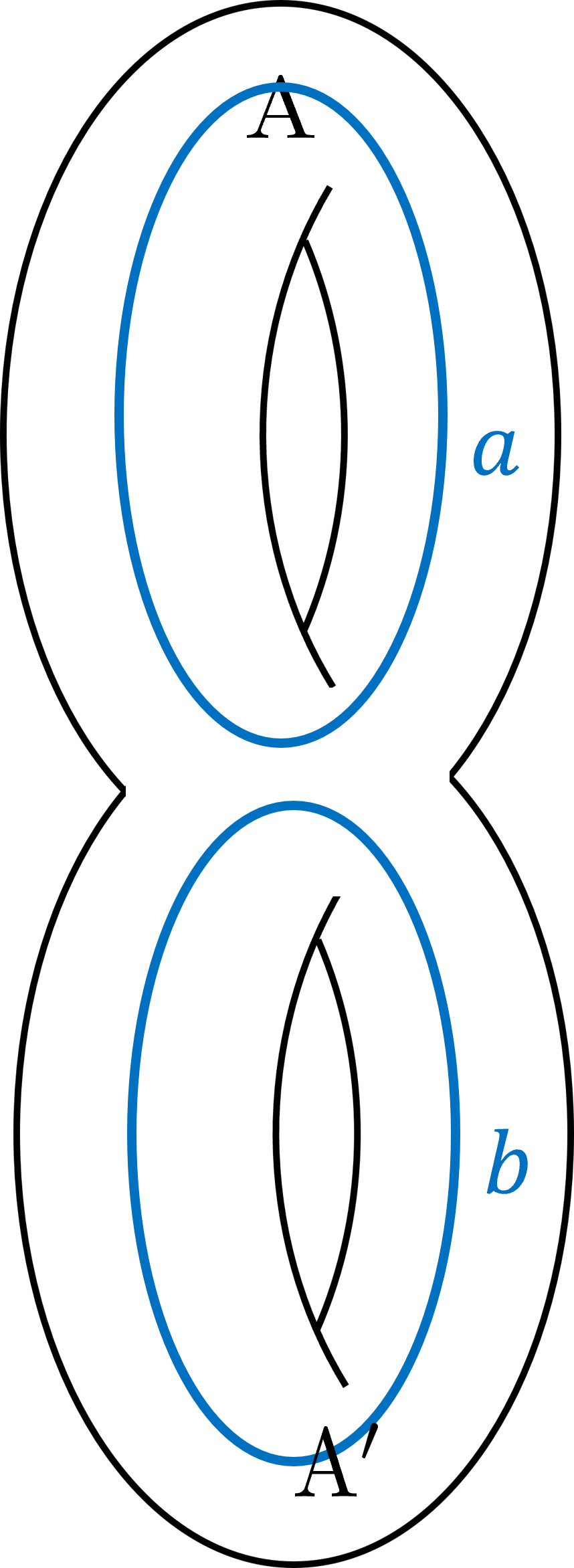}}\hspace{0.1cm}  
    = \sum_{a,b} \frac{d_a}{D} \frac{d_b}{D}\hspace{0.3cm}
    \adjustbox{valign = c}{\includegraphics[height = 5cm]{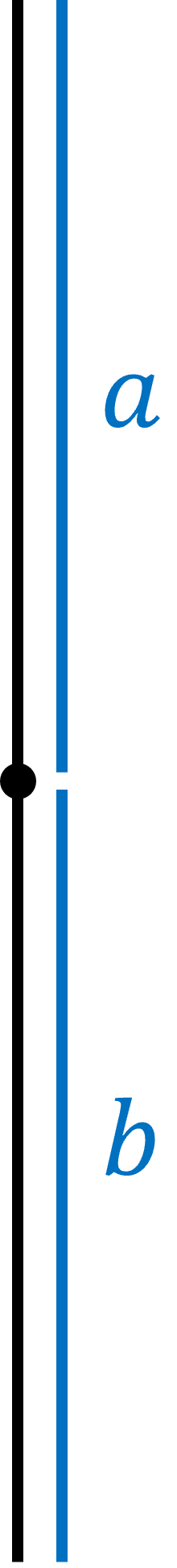}}\hspace{0.1cm},
    \label{Eq46}
\end{equation}
where $d_a = d_{\Bar{a}} = S_{0a} D$ is the quantum dimension of anyon $a$ and $D = \sqrt{\sum_{a} d_a^2}$ is the total quantum dimension. 
For simplicity, in the right-hand side of Eq.~(\ref{Eq46}), we use a single black line to indicate a copy of $D^2 \times S^1$, we use the dot to indicate that they are being connected summed and we use blue lines to indicate the anyon loops circling the corresponding non-contractible loops of $D^2 \times S^1$. Using this notation, we have
\begin{equation}\label{Tr_rho^n}
    \mathrm{Tr}{(\rho_A}^n) = \sum_{a_1,b_1} \sum_{a_2,b_2} \dots \sum_{a_n,b_n} \Big(\prod_{k=1}^n \frac{d_{a_k} d_{b_k}}{D^2} \Big)\hspace{0.2cm}
    \adjustbox{valign = c}{\includegraphics[height = 5cm]{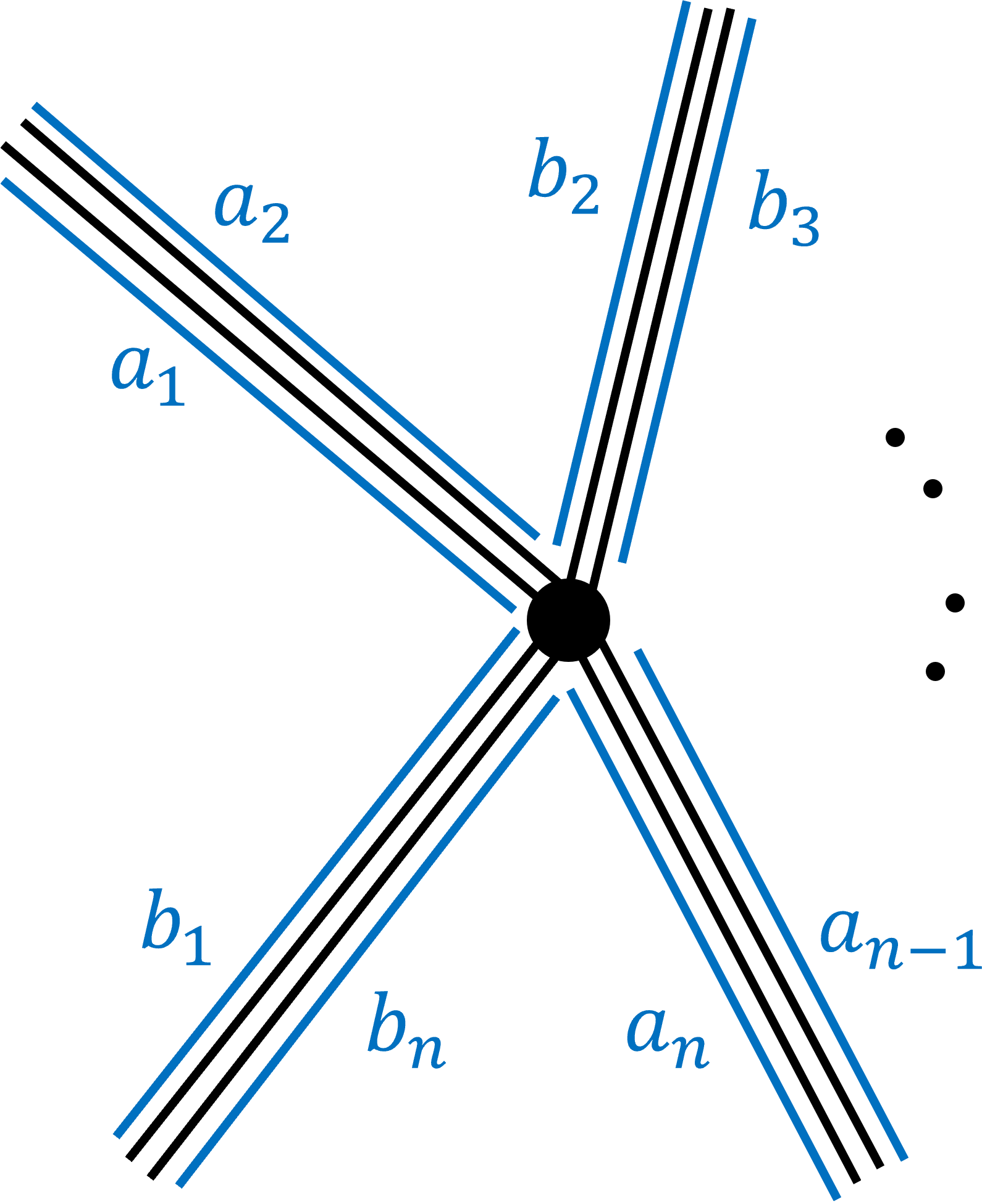}}\hspace{0.1cm},
\end{equation}
where the double lines indicates that the two copies of $D^2 \times S^1$ are glued together along their torus boundary to form a $S^2 \times S^1$, and the giant dot indicates that these $S^2 \times S^1$ are connected summed together. We should also mentioned that all the lines are all in the outside basis even if we do not put them in the vertical manner. To evaluate the partition function, we apply the method of surgery again to cut out all the separate $S^2 \times S^1$'s with Wilson lines by inserting $S^3$'s. We obtain
\begin{equation}
    \mathrm{Tr}{(\rho_A}^n) = \sum_{a_1,b_1} \sum_{a_2,b_2} \dots \sum_{a_n,b_n} \prod_{k=1}^n \frac{d_{a_k} d_{b_k}}{D^2}\hspace{0.1cm}
    \adjustbox{valign = c}{\includegraphics[height = 2.5cm]{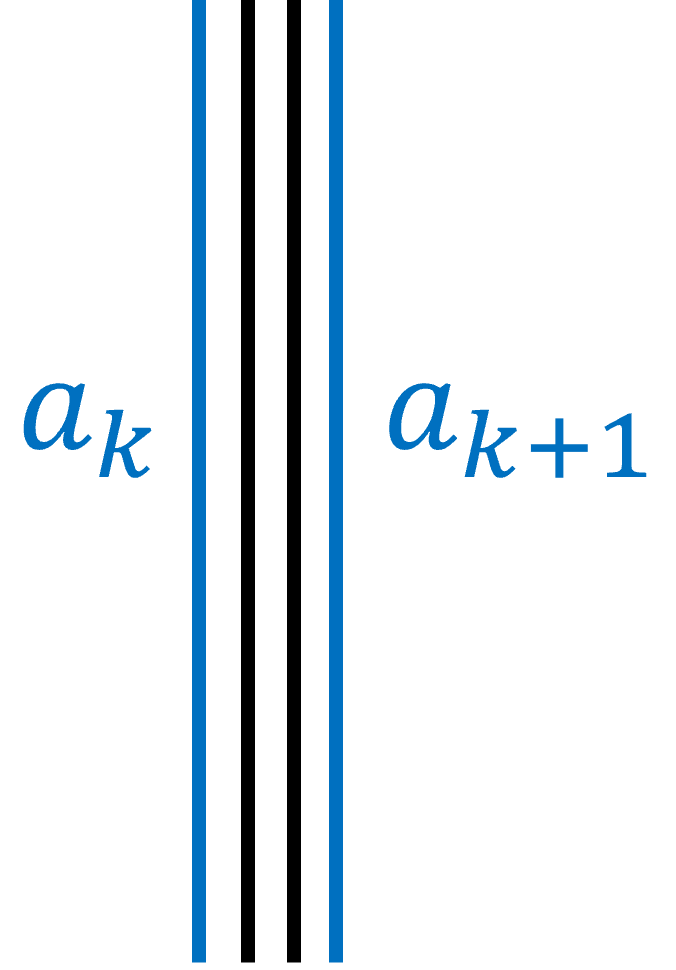}}\hspace{0.1cm}
    \adjustbox{valign = c}{\includegraphics[height = 2.5cm]{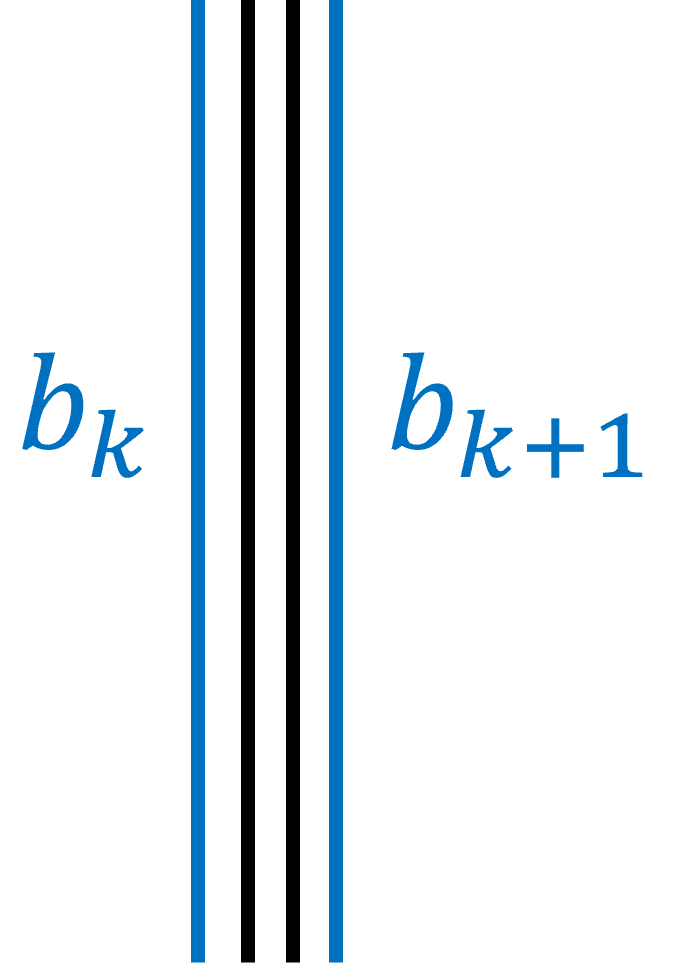}}\hspace{0.1cm}
    \frac{1}{Z(S^3)^{n-1}}.
\end{equation}
Since the partition function of each $S^2 \times S^1$ with Wilson lines $\Bar{a}, b$ is given by $\delta_{ab}$, and the partition function of $S^3$ is simply $D^{-1}$, we have
\begin{equation}
    \mathrm{Tr}{(\rho_A}^n) = \Big( \sum_a \frac{d_a^2 }{D^2} \Big)^n D^{n-1} = D^{n-1}.
\end{equation}
As expected, we obtain the same result as using the inside basis. This serves as a consistency check of our method.

\subsection{Two \texorpdfstring{$S^1$}{} interfaces related by modular \texorpdfstring{$S$}{} transformation}\label{subsection4.2}
Next, we consider the case where the two sub-regions meet at two $S^1$ interfaces. If both $S^1$ interfaces are contractible, then the result will be similar to the case of a single interface. If only one $S^1$ is contractible, then the bipartition is ill-defined. The next possible bipartition is when the two interfaces are given by the same torus knots. In this subsection, we will consider the case where the torus knots are simply given by the longitudes and meridians.

\subsubsection{Two \texorpdfstring{$S^1$}{} interfaces being the longitudes of the torus}
Let us consider the vacuum state in the solid torus and with the bipartition given by two longitudes which can be viewed as cutting the doughnut horizontally
\begin{equation}\label{longitude}
    \ket{0} = \hspace{0.2cm}
    \adjustbox{valign = c}{\includegraphics[height = 3cm]{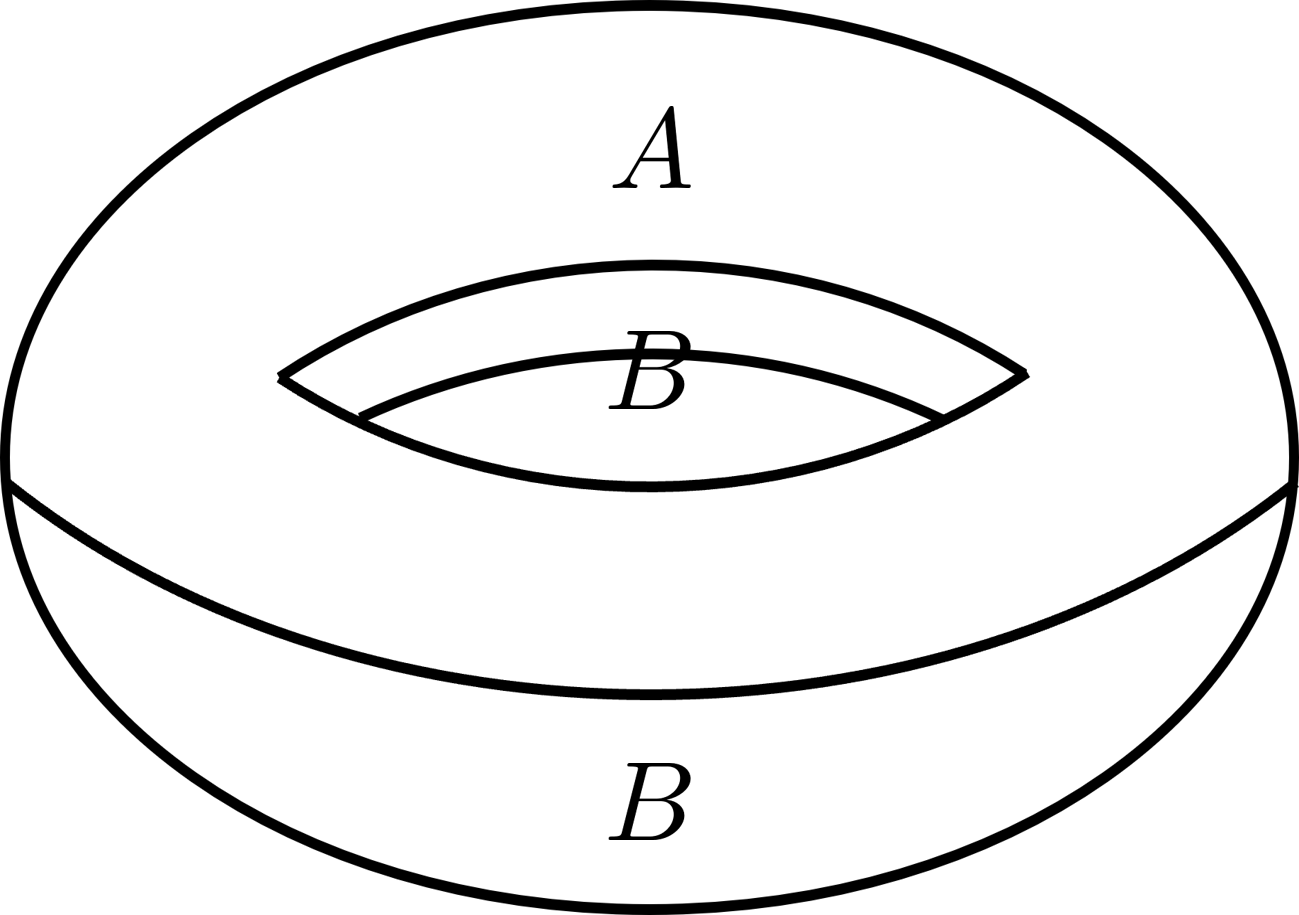}}\hspace{0.1cm}.
\end{equation}
In this case, the reduced density matrix after tracing out $B$ is given by
\begin{equation}
    \rho_A = \hspace{0.2cm}
    \adjustbox{valign = c}{\includegraphics[height = 3cm]{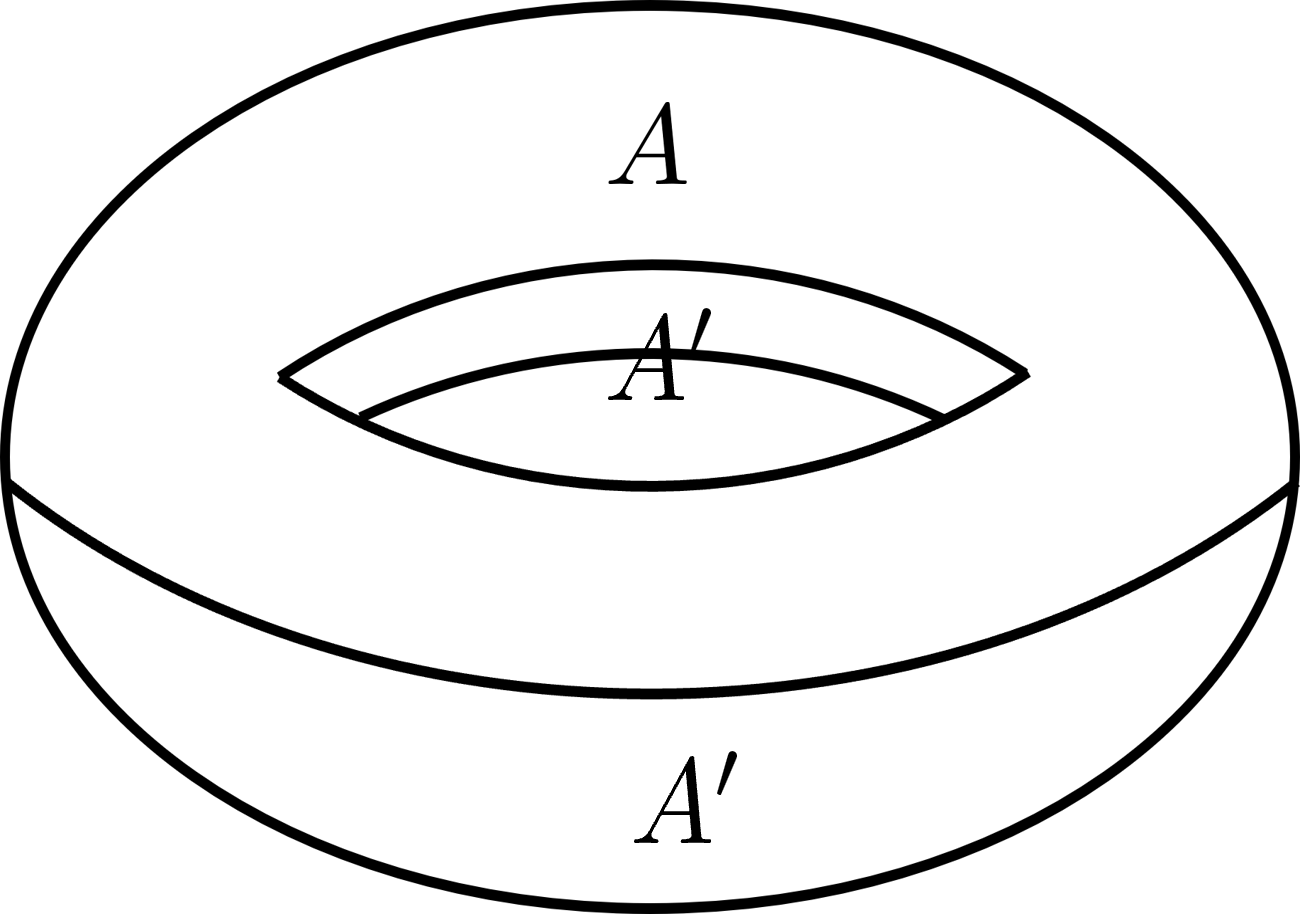}}\hspace{0.1cm}.
\end{equation}
We have 
\begin{equation}
    \mathrm{Tr}({\rho_A}^n) = Z(S^2 \times S^1) = 1.
\end{equation}
On the other hand, the calculation in the outside basis is rather non-trivial. 
We have
\begin{equation}
    \ket{0} = \hspace{0.2cm}
    \adjustbox{valign = c}{\includegraphics[height = 3cm]{donut_AB_3.png}}\hspace{0.1cm}
    = \sum_{a} S_{0a} \hspace{0.2cm}
    \adjustbox{valign = c}{\includegraphics[height = 5cm]{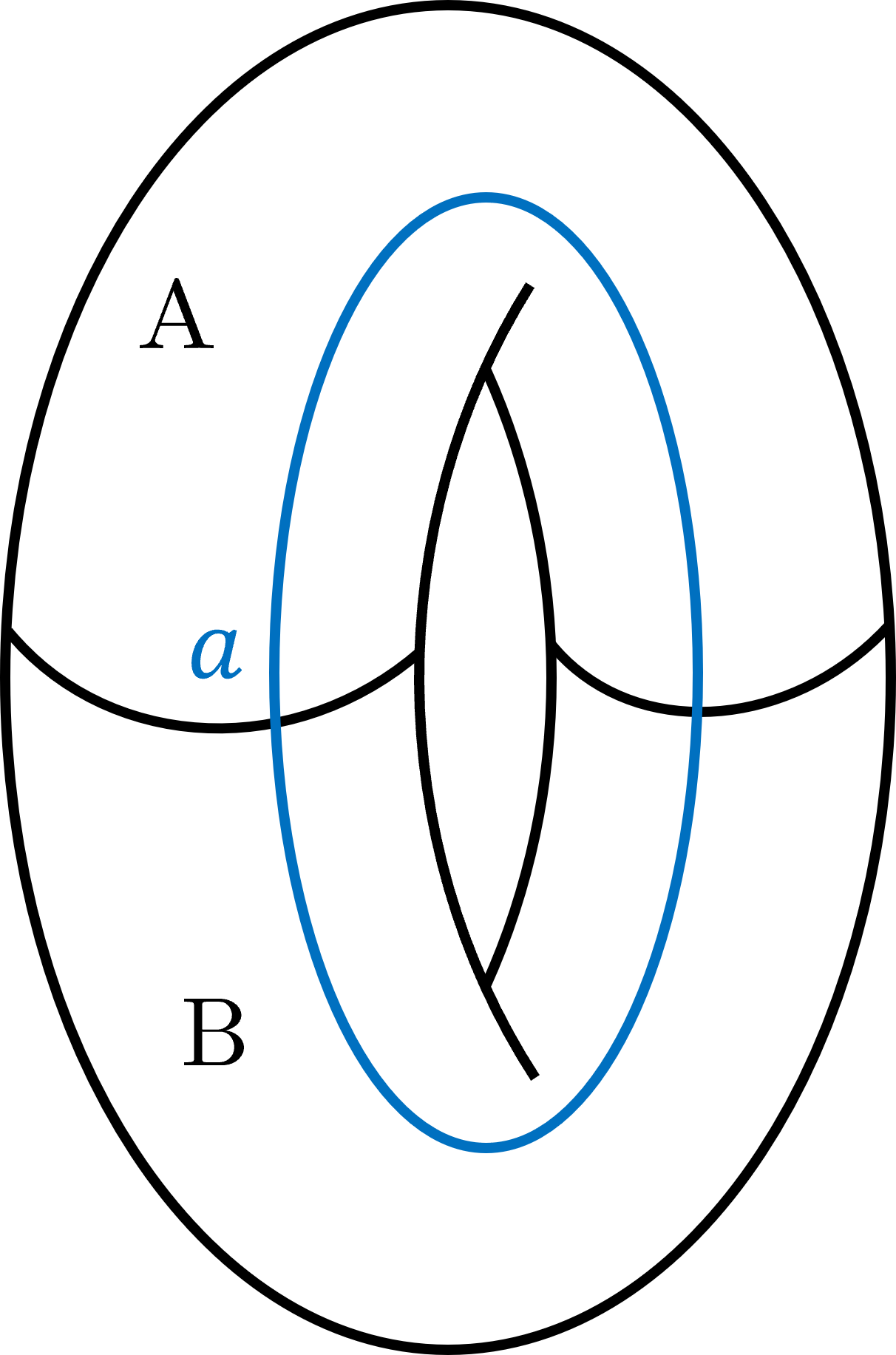}}\hspace{0.1cm},
\end{equation}
where the bipartition also transforms with respect to the rewiring of sites as discussed in Sec.~\ref{section:coordinate}.
We obtain the reduced density matrix $\rho_A$ by gluing $B$ to $B'$,
\begin{equation}
    \rho_A = \sum_{a,b} S_{0a} S_{0 \Bar{b}}\hspace{0.2cm}
    \adjustbox{valign = c}{\includegraphics[height = 4cm]{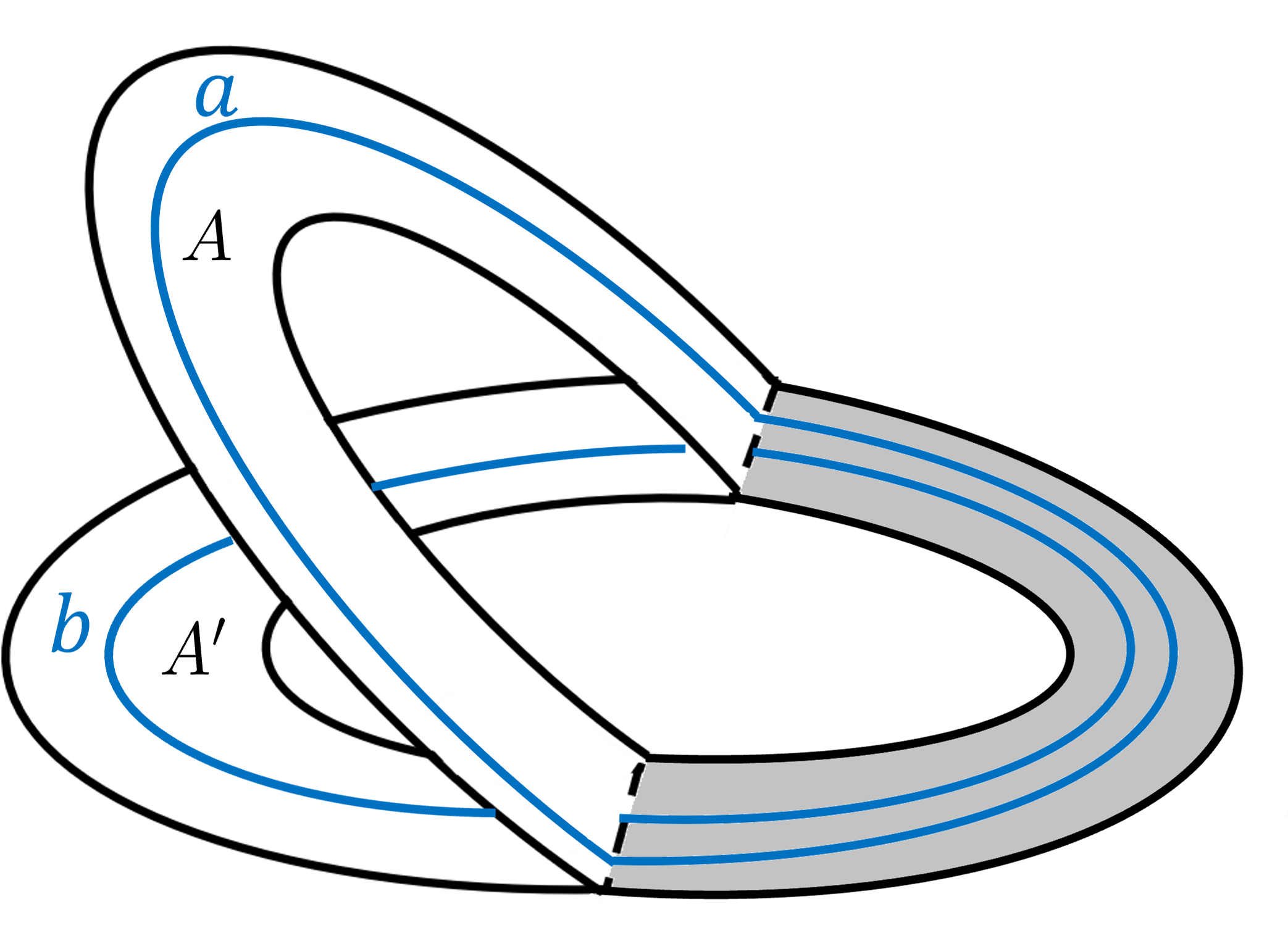}}\hspace{-6mm}
    = \sum_{a,b} S_{0a} S_{0 b}\hspace{0.2cm}
    \adjustbox{valign = c}{\includegraphics[height = 3cm]{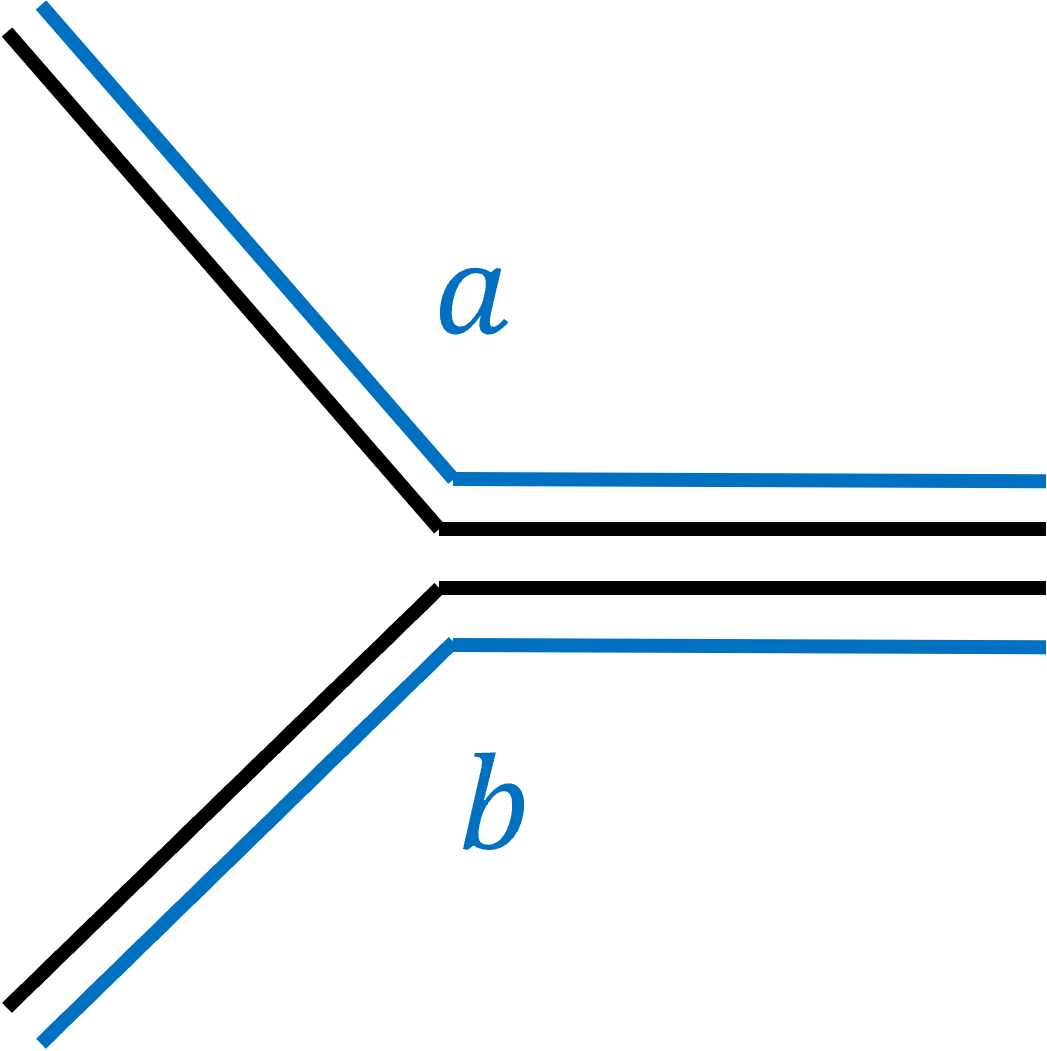}}\hspace{0.1cm}.
    \label{Eq:A}
\end{equation}

Here the graph in the left-hand side of Eq.~(\ref{Eq:A}), we flatten our solid torus to be an annulus for simplicity. The glued region indicated by color gray is $S^2 \times D^1$ with two copies of $S^2$ boundaries. Also, each half solid torus ($A$ and $A'$) has boundary being two copies of $D^2$. Therefore, two $D^2$ of separate $A$ and $A'$ then be glued to form $S^2$ which matches each side of the boundary of the $S^2 \times D^1$. The right-hand side of Eq.~(\ref{Eq:A}) is the view at the angle parallel to the radial direction of the circles (side view). For example,
\begin{equation}
    \mathrm{Tr}{\rho_A}^2 = \hspace{0.1cm}
    \adjustbox{valign = c}{\includegraphics[height = 5cm]{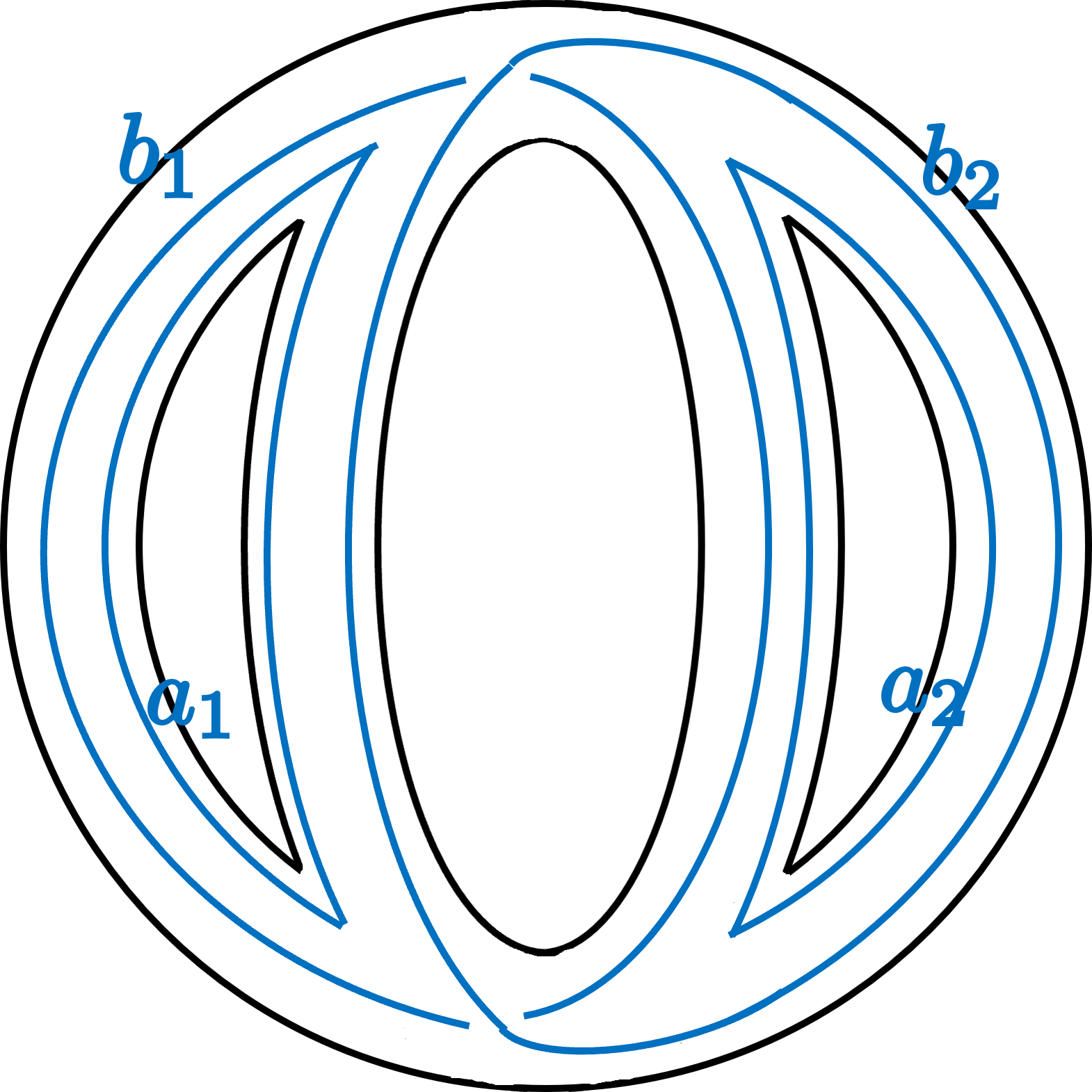}}\hspace{0.1cm}
    = \hspace{0.1cm}
    \adjustbox{valign = c}{\includegraphics[height = 3cm]{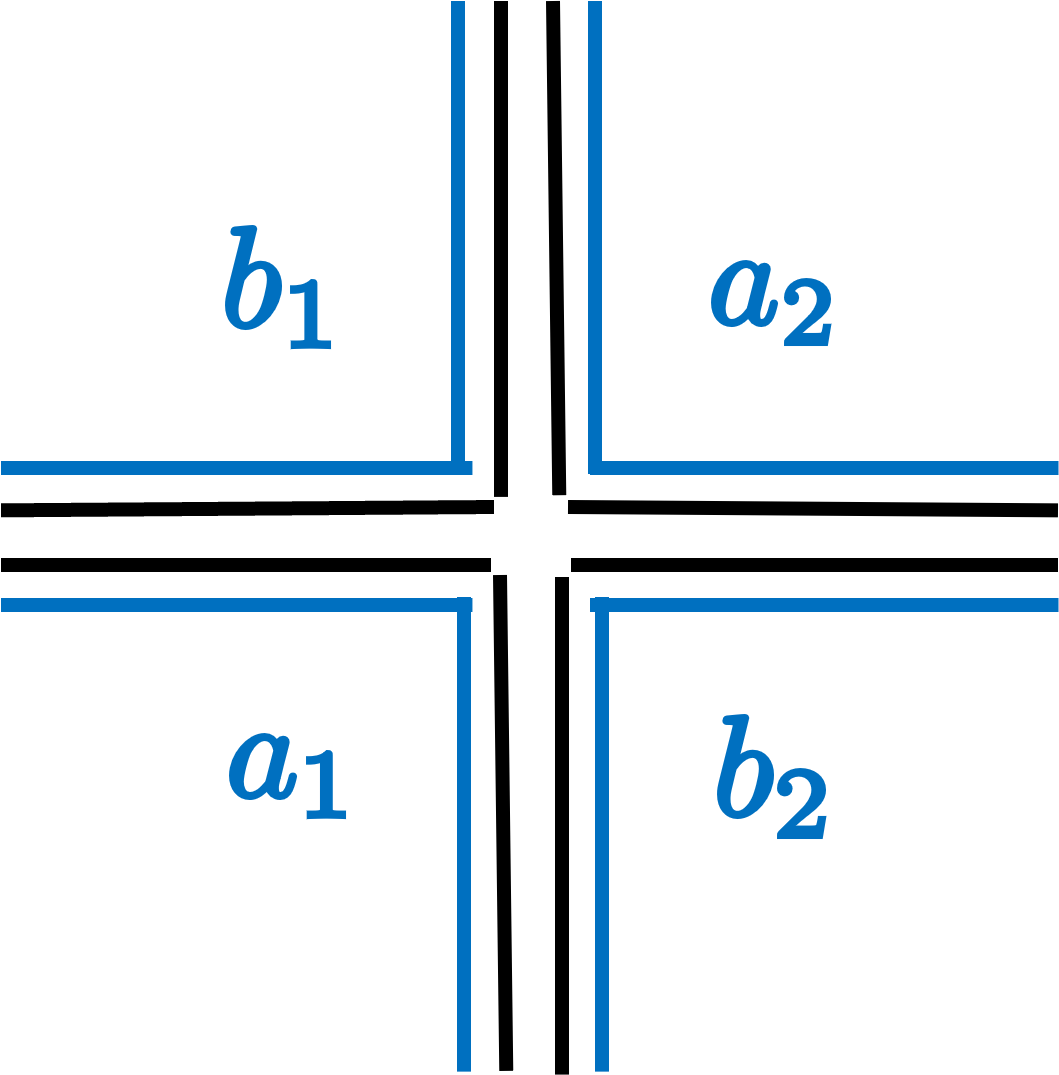}}\hspace{0.1cm}.
    \label{Eq:B}
\end{equation}
With this notation, we have
\begin{equation}
    \mathrm{Tr}{(\rho_A}^n) = \sum_{\substack{a_1,\dots ,a_n\\ b_1, \dots ,b_n}} \big( \prod_{i=1}^n \frac{d_{a_i}}{D} \frac{d_{b_i}}{D} \big) \hspace{0.1cm}.
    \adjustbox{valign = c}{\includegraphics[height = 5cm]{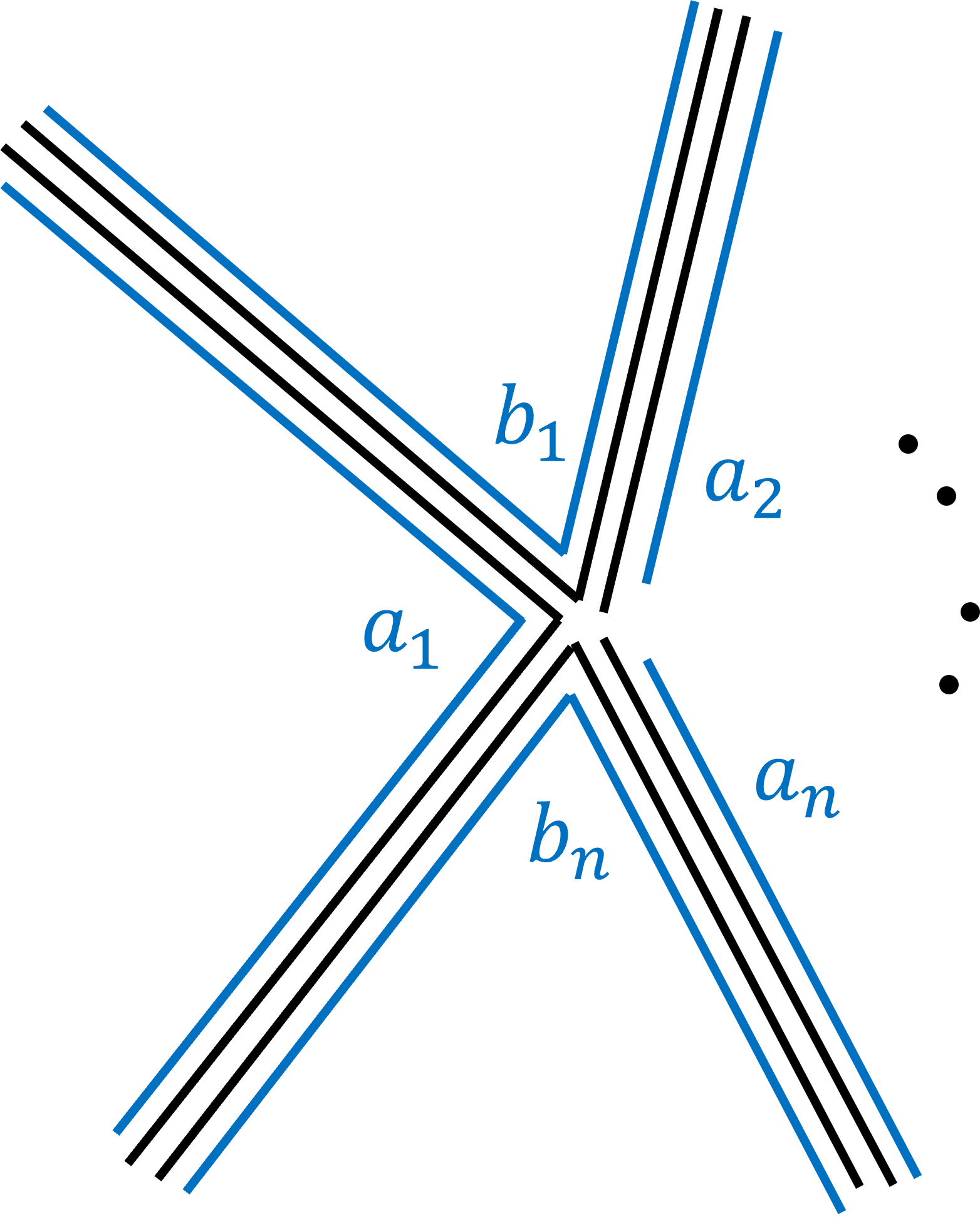}}\hspace{0.1cm}.
\end{equation}
We should note that each double line now represents $S^2 \times D^1$ which is different from we have seen in Eq.~(\ref{Tr_rho^n}). Each $S^2 \times D^1$ has two disconnected $S^2$ boundaries and each $S^2$ boundary is separately connected summed to other corresponding $S^2$ boundaries. Next, we flatten the configuration by pushing all the handles ($S^2 \times D^1$) in to the same plane in the following fashion
\begin{equation}\label{flatten}
    \adjustbox{valign = c}{\includegraphics[height = 1.8cm]{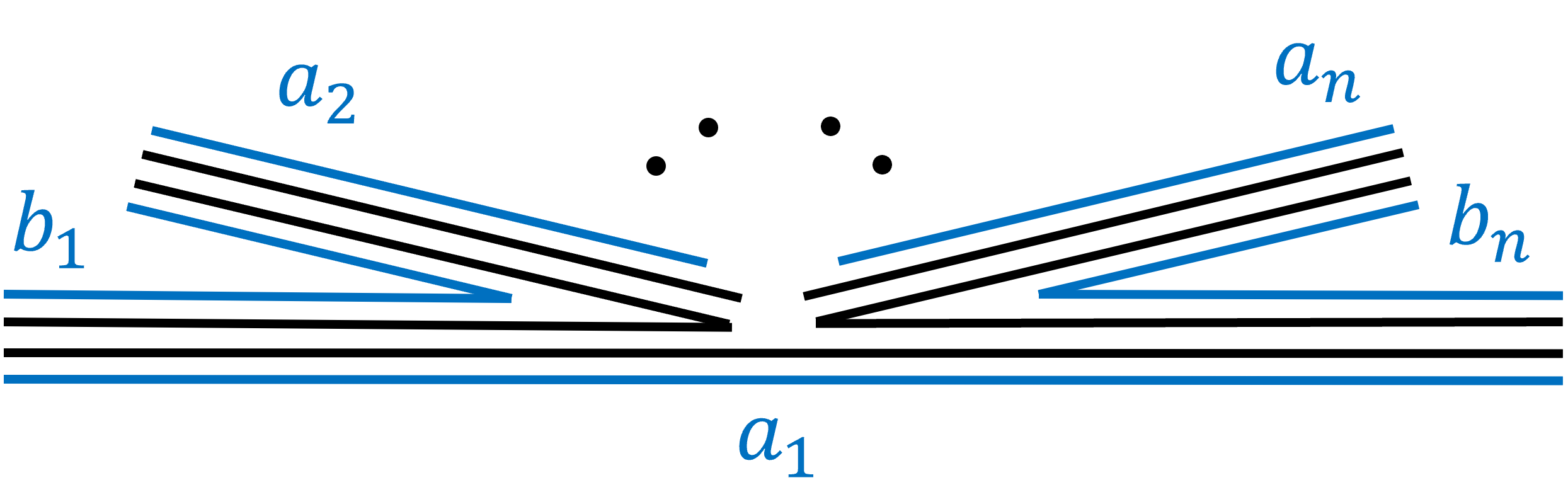}}\hspace{0.1cm}
    = \hspace{0.1cm}
    \adjustbox{valign = c}{\includegraphics[height = 2cm]{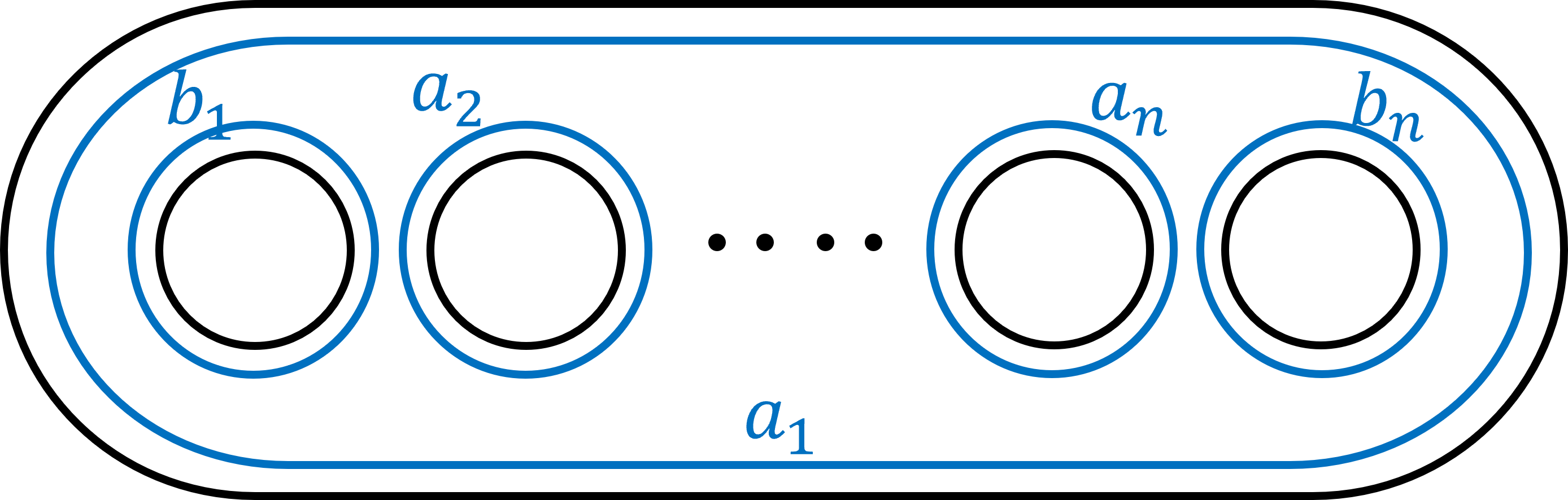}}\hspace{0.1cm}.
\end{equation}
The right-land side of Eq.~(\ref{flatten}) is a connected sum of $S^2 \times S^1$, where each hole represent a copy of $S^2 \times S^1$. The anyon lines are given from left to right in the order $b_1, a_2, ..., b_{k-1}, a_k, ...b_{n-1}, a_n, b_n$ running in the same direction and $a_1$ winds all the holes. 
For example, one can refer to the left-hand side of Eq.~(\ref{Eq:B}) and deform it to the right-land side of Eq.~(\ref{flatten}) with three holes.
We then apply the method of surgery by inserting $2n-2$ copies of $S^3$ with $a_1$ anyon loop as in Eq.~(\ref{eqn:surgery_line}). Therefore, we can obtain that
\begin{equation}
    \mathrm{Tr}{(\rho_A}^n) = \sum_{\substack{a_1,\dots ,a_n\\ b_1, \dots ,b_n}} \prod_{i=1}^n \frac{d_{a_i}}{D} \frac{d_{b_i}}{D} \delta_{\bar{a_1} a_i} \delta_{\Bar{a_1} b_i} \Big(\frac{D}{d_{a_1}} \Big)^{2n-2} = \sum_{a_1} \frac{d_{a_1}^2}{D^2} = 1
\end{equation}
Although the intermediate steps are less trivial, the final results using the outside basis and inside basis match.

\subsubsection{Two \texorpdfstring{$S^1$}{} interfaces being the meridians of the torus}
Next, let us consider the vacuum state whose bipartition contain two meridians of the torus
\begin{equation}\label{meridian}
    \ket{0} = \hspace{0.2cm}
    \adjustbox{valign = c}{\includegraphics[height = 2.5cm]{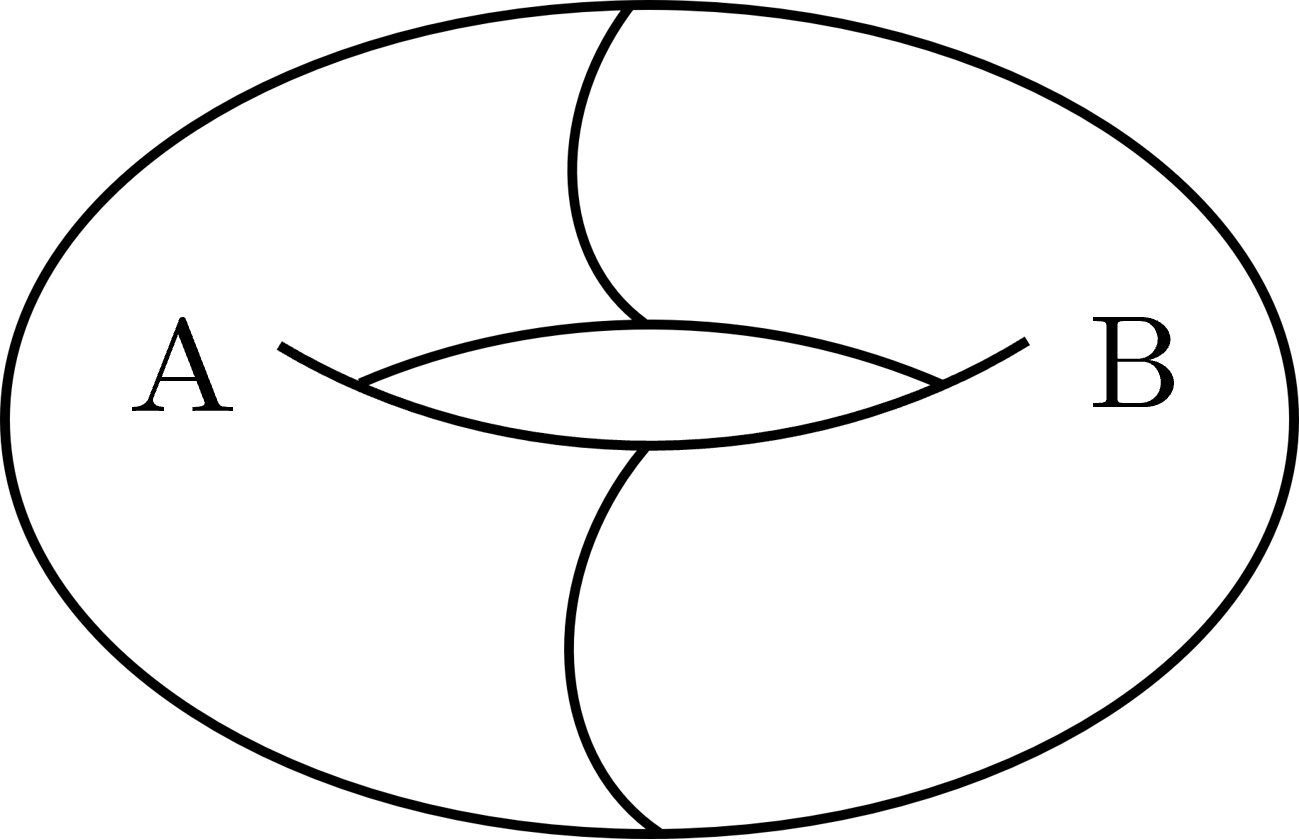}}\hspace{0.1cm}.
\end{equation}
In this case, we have the reduced density matrix
\begin{equation}
    \rho_A = \hspace{0.2cm}
    \adjustbox{valign = c}{\includegraphics[height = 4cm]{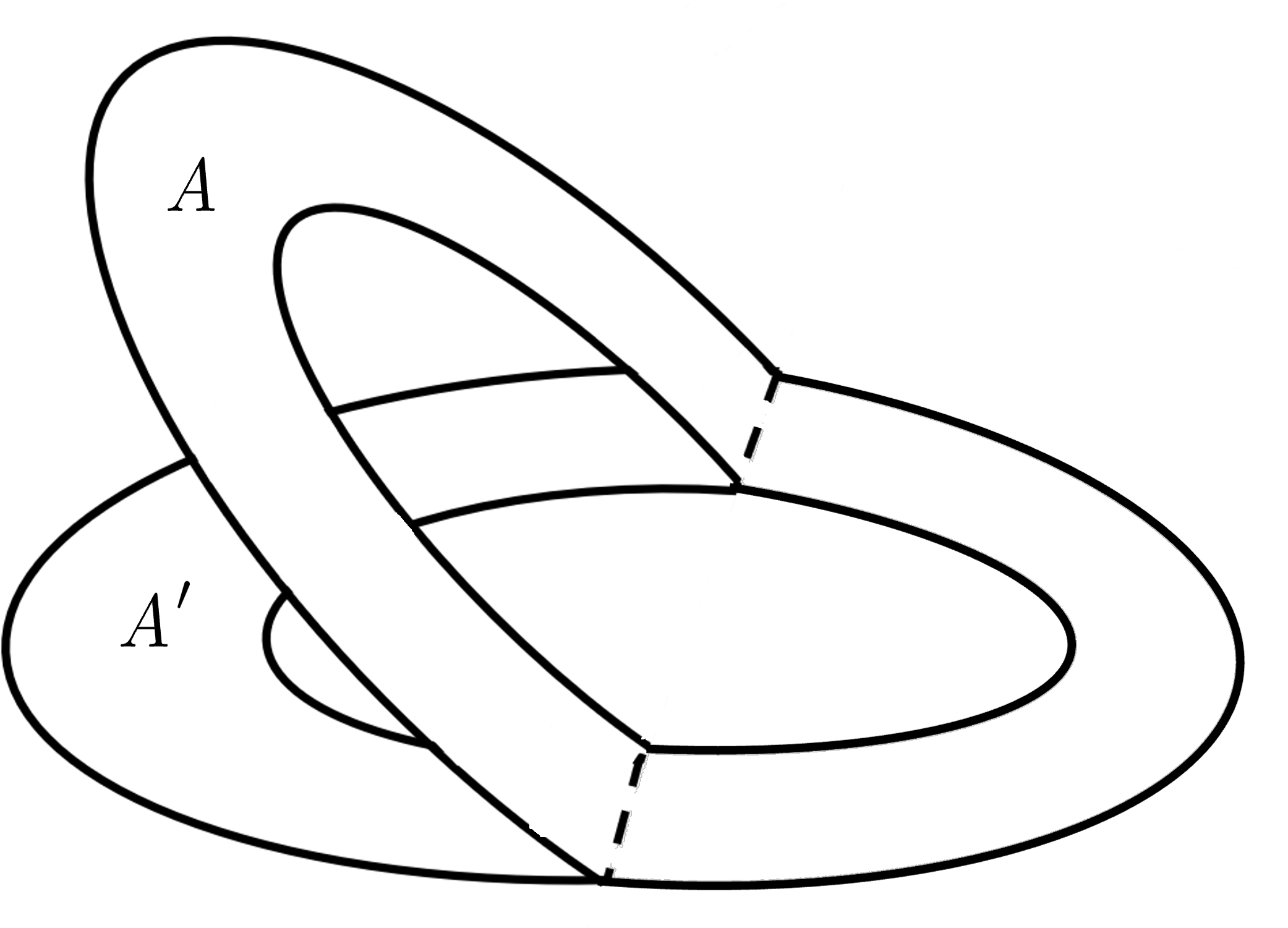}}\hspace{0.1cm},
\end{equation}
where we flatten the tori similar to previous section. Following the same procedure, we obtain the same configuration as equation Eq.~(\ref{flatten}) but without the anyon lines. The final result is a connected sum of $2n-1$ copies of $S^2 \times S^1$. Therefore, we apply the method of surgery by inserting $2n-2$ copies of $S^3$ and get
\begin{equation}\label{tr_rho^n_2}
    \mathrm{Tr}{(\rho_A}^n ) = \frac{Z(S^2 \times S^1)^{2n-1}}{Z(S^3)^{2n-2}} = D^{2n-2}.
\end{equation}
Now, we do the same calculation but using the outside basis instead. The change of basis is
\begin{equation}
    \ket{0} = \hspace{0.2cm}
    \adjustbox{valign = c}{\includegraphics[height = 2.5cm]{donut_AB_2.png}}\hspace{0.1cm}
    = \sum_{a} S_{0a} \hspace{0.2cm}
    \adjustbox{valign = c}{\includegraphics[height = 5cm]{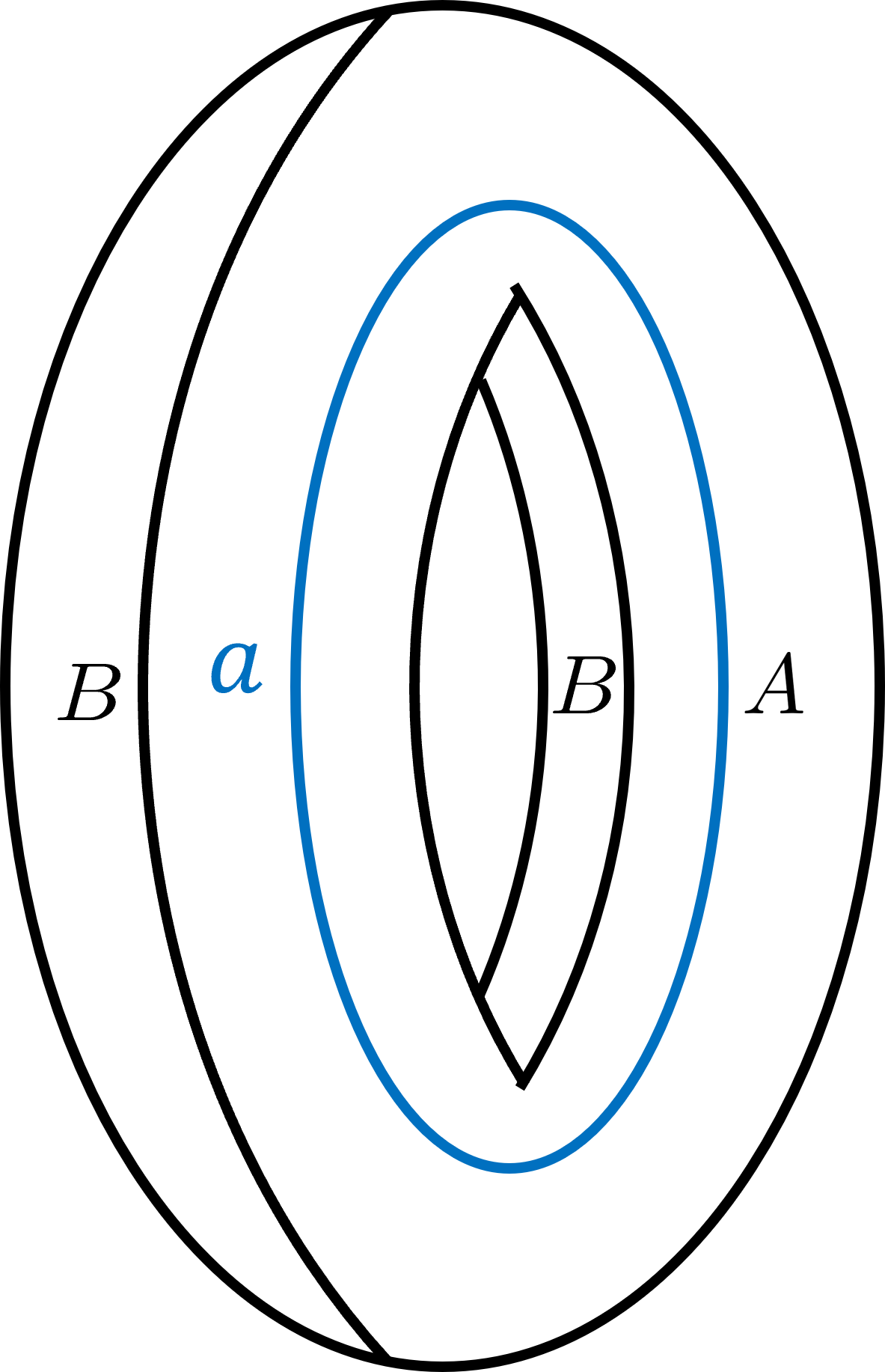}}\hspace{0.1cm}.
\end{equation}
We have
\begin{equation}
    \rho_A = \sum_{a,b} S_{0a} S_{0b}\hspace{0.2cm}
    \adjustbox{valign = c}{\includegraphics[height = 5cm]{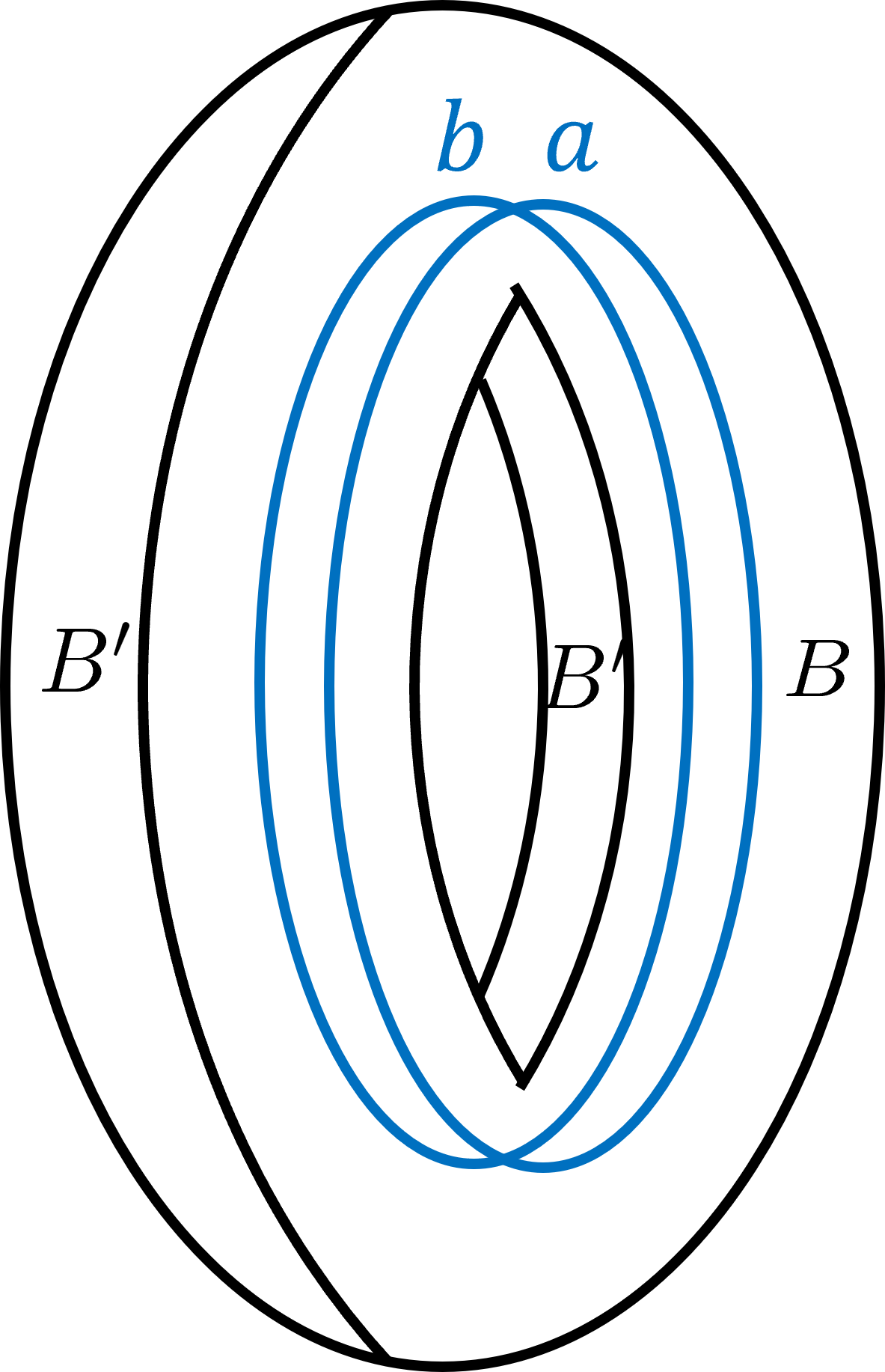}}\hspace{0.1cm},
\end{equation}
and
\begin{equation}
    \mathrm{Tr}{(\rho_A}^n) = \sum_{a_1,b_1} \sum_{a_2,b_2} \dots \sum_{a_n,b_n} \prod_{k=1}^n \frac{d_{a_k} d_{b_k}}{D^2}\hspace{0.1cm}
    \adjustbox{valign = c}{\includegraphics[height = 5cm]{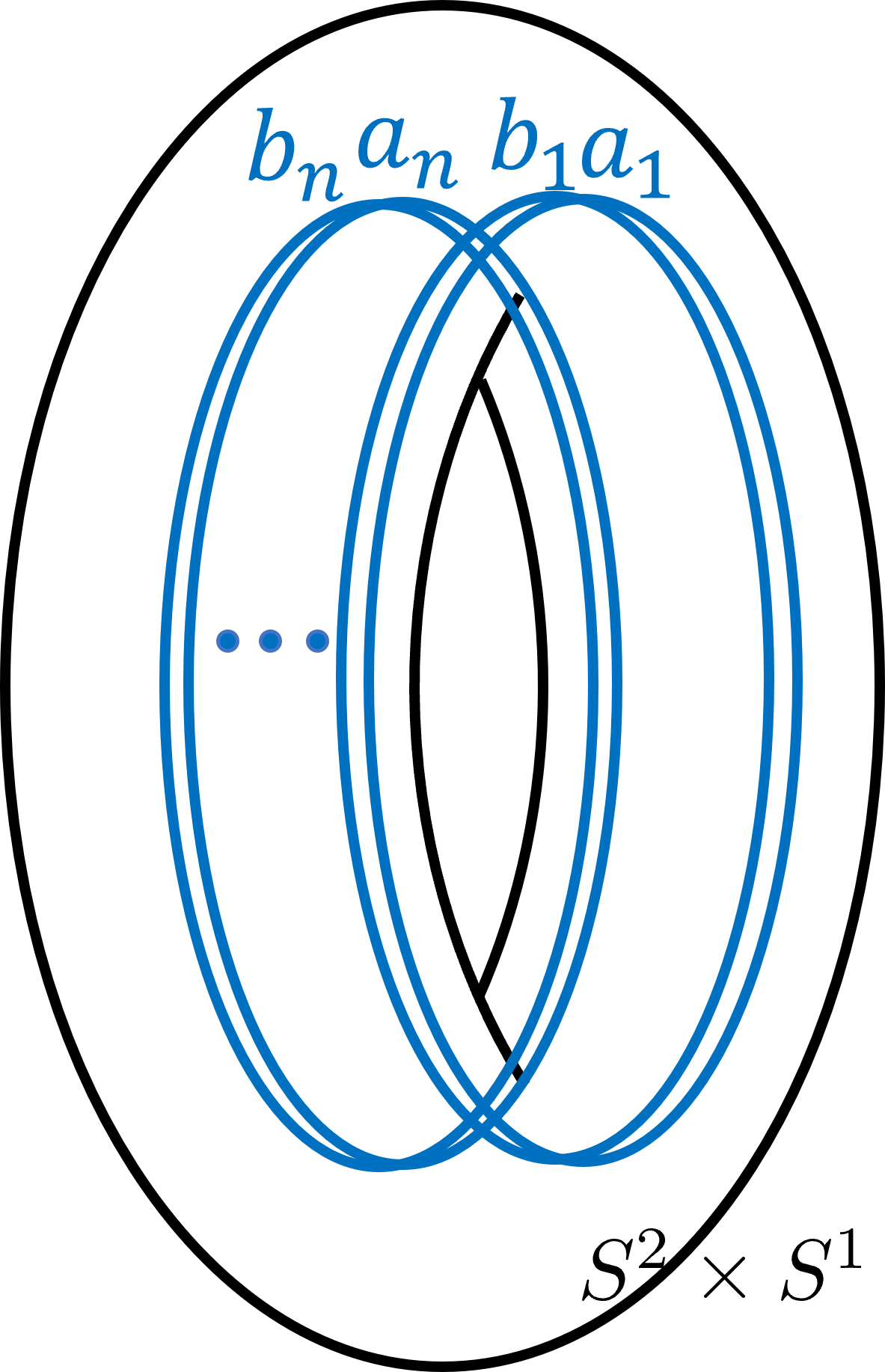}}\hspace{0.1cm}.
\end{equation}
The partition function of $S^2 \times S^1$ with anyon lines $a_k$, $b_k$ in its non-contractible loop is given by $\mathrm{dim} V_{a_1,b_1, \dots a_n,b_n}$.
Equating this to Eq.~(\ref{tr_rho^n_2}), we obtain
\begin{equation}\label{main}
    \sum_{a_1,b_1} \sum_{a_2,b_2} \dots \sum_{a_n,b_n}  d_{a_1} d_{b_1} \dots ,d_{a_n} d_{b_n} \mathrm{dim} V_{a_1,b_1, \dots a_n,b_n} = D^{4n-2}.
\end{equation}

This formula tells us that summing over the product of quantum dimensions of all the possible arrangements of $2n$ anyons weighted by the fusion-tree dimensions $\mathrm{dim} V_{a_1,b_1, \dots, a_n,b_n}$ equals to the total quantum dimension $D^{4n-2}$.
Eq.~(\ref{main}) reminds us of the Verlinde formula~\cite{Ver, genVer}, which also relates the quantum dimensions to the modular data.
For example, in the case where $n=1$, $\mathrm{dim} V_{a_1,b_1} = \delta_{a_1,b_1}$, and thus we have the definition of quantum dimension $\sum_a d_a^2 = D^2$.
Although not quite obvious, one can also derive Eq.~(\ref{main}) using the Verlinde formula, which we will discuss in appendix~\ref{appendixA}.

\subsection{General Torus-knot bipartitions}\label{subsection4.3}

\begin{figure}
    \centering
    \includegraphics[height=4cm]{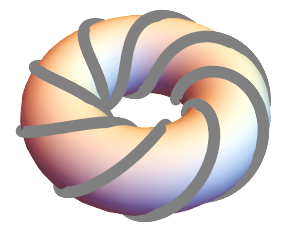}
    \includegraphics[height=5cm]{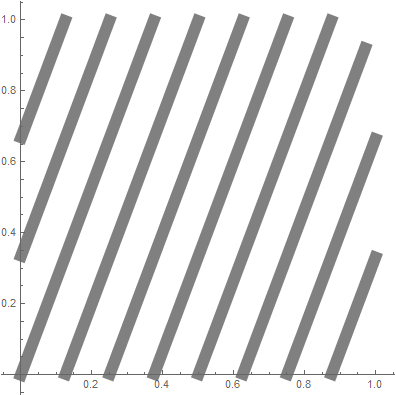}
    \caption{Torus knot $K(8,3)$ embedded in $\mathbb{R}^3$ and on its fundamental domain.}
    \label{torusknot}
\end{figure}

Having established the consistency of the calculation of entanglement quantities using different coordinates, we now compute the entanglement quantities for more complicated configurations. In particular, we consider bipartitions with two $S^1$ interfaces, both being torus knots. We will refer to such bipartitions as torus-knot bipartitions throughout this paper. To visualize a torus-knot bipartition, one can imagine a single torus knot on a torus, as in Fig.~\ref{torusknot}, and widen the line to become a ribbon that defines the subregion.

A general torus knot can be written as $K(k_a,k_b)$, where $k_a$ and $k_b$ are integers with $\mathrm{gcd}(k_a, k_b) = 1$. Let $K(1,0)$ and $K(0,1)$ denote the meridians and the longitudes, then $K(k_a,k_b)$ is a torus knot that winds $k_a$ and $k_b$ times in the meridian and longitude direction, respectively. For example, Fig. \ref{torusknot} shows a torus knot $K(8,3)$.
Suppose we have a bipartition $T^2 = A \amalg B$, with interfaces $\partial A = \partial B$ composed of two $S^1$ interfaces, then the two $S^1$ interfaces should be the same torus knot to avoid intersecting with each other. Therefore, we can uniquely determine a torus-knot bipartition by the type of torus knots its interfaces are made of. In previous sections, our focus has been on cases where the bipartition consists of two $K(1,0)$ interfaces or two $K(0,1)$ interfaces, which we refer to as untwisted bipartitions. Specifically, for the bipartition given by two $K(1,0)$ interfaces, we term it the canonical bipartition. In this section, we will compute the TEEs for  torus-knot bipartitions by transforming them back to the canonical bipartition.\par
Let us consider a state $\ket{0}$, which is generated by a solid torus with no Wilson lines inserted, and a torus-knot bipartition on its boundary with the interfaces being two $K(k_a, k_b)$ torus knots with $\mathrm{gcd}(k_a, k_b) = 1$. By applying modular $S$ and $T$ transformations, one can transform $K(k_a, k_b)$ into $K(k_b, k_a)$ and $K(k_a, k_b + k_a)$, respectively. Therefore, using the Euclidean algorithm, one can transform the torus knot into the meridian $K(1,0)$ by a series of modular $S$ and $T$ transformations.
\begin{figure}
    \centering
    \includegraphics[height=6cm]{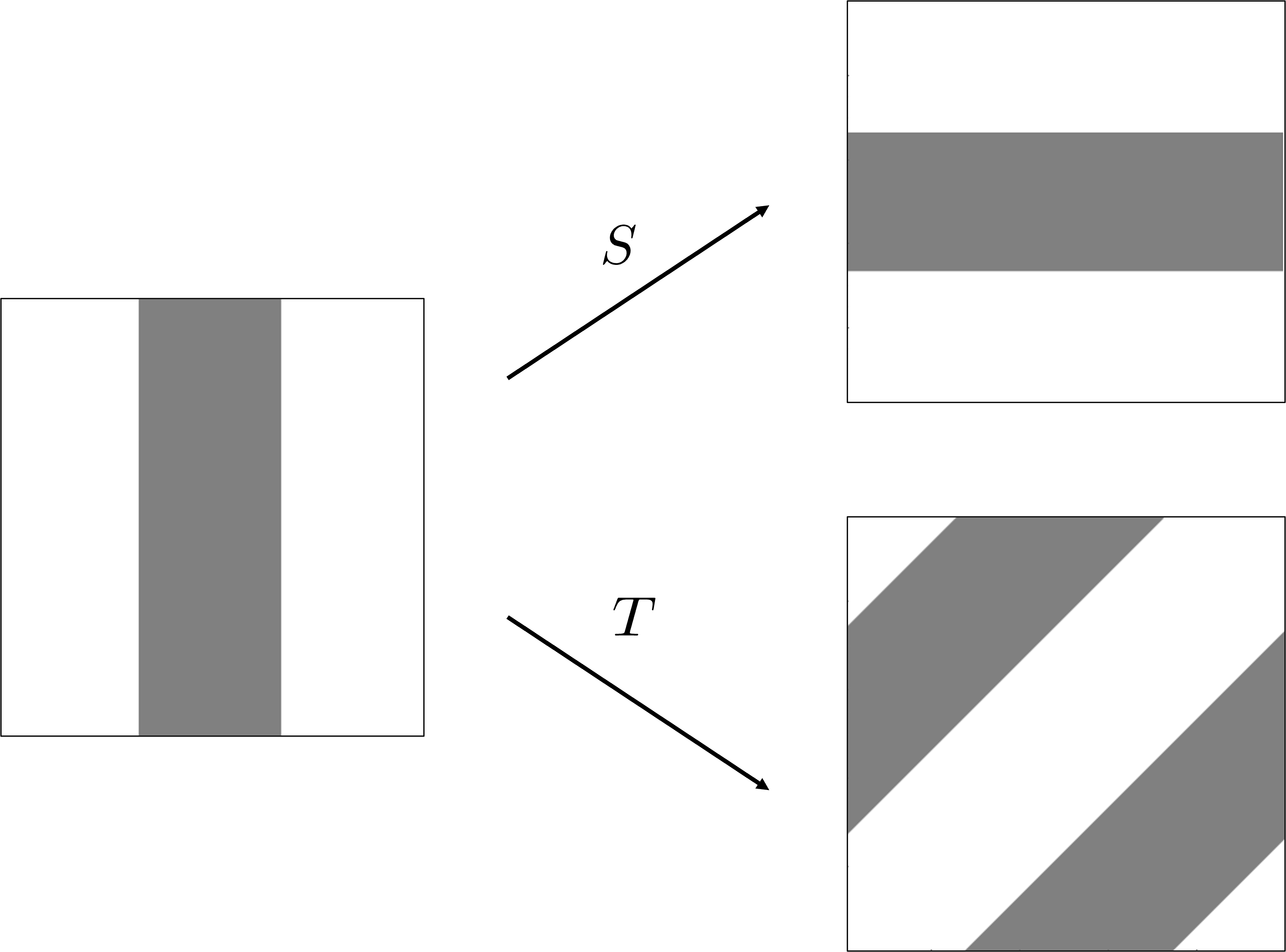}
    \caption{The modular $S$ and $T$ transformation generate the whole modular group $SL(2,\mathbb{Z})$.}
    \label{modulartrans}
\end{figure}
That is, all the possible torus-knot bipartitions can be mapped back to the canonical bipartition consist of meridian interfaces by some $\mathcal{O} \in SL(2,\mathbb{Z})$. For example, $K(8,3)$ can be turned into the canonical bipartition by the sequentially application of $ST^2 S^{-1}, T, ST^2 S^{-1}, S$ on the torus. The combined transformation is then given by $\mathcal{O} = S ST^2 S^{-1} T ST^2 S^{-1} = (ST)^{-1}$. With the expression of $\mathcal{O}$, we rewrite 
\begin{equation}\label{twist}
    \ket{0} = \mathcal{O} \mathcal{O}^{-1} \hspace{0.2cm}
    \adjustbox{valign = c}{\includegraphics[height = 3cm]{torus_eg_1}}\hspace{0.1cm}
    =  \sum_{a} \mathcal{O}_{0a} \hspace{0.2cm}
    \adjustbox{valign = c}{\includegraphics[height = 3cm]{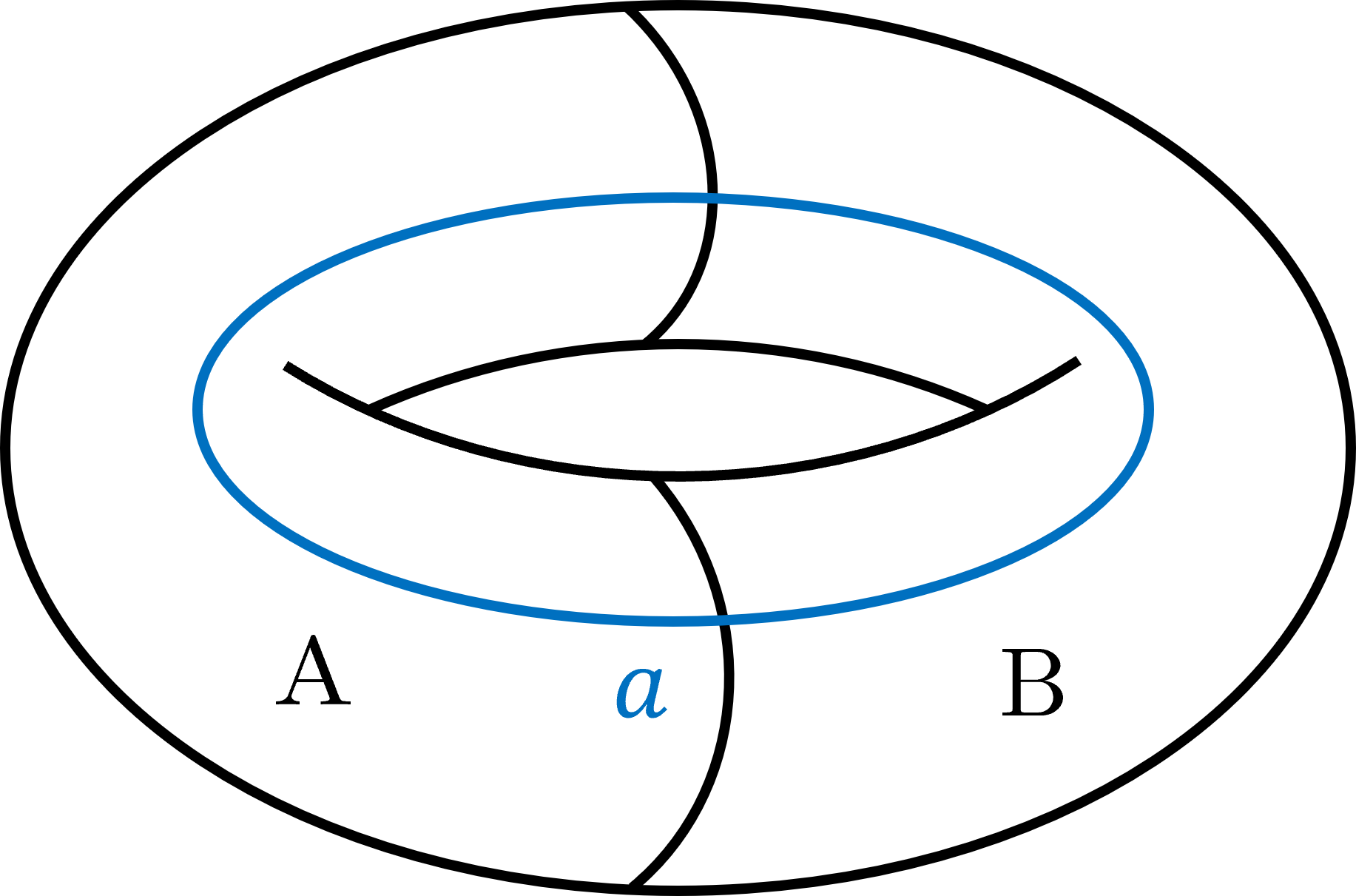}}\hspace{0.1cm}.
\end{equation}
The gray line in the left figure of Eq.~(\ref{twist}) indicate the subregion $A$ and the solid torus has no Wilson lines insert. In the right figure of Eq.~(\ref{twist}), the coordinate grid is twisted compared to the left figure. However, we can perform exactly the same replica method in the new coordinate system as long as all the copies are in the same coordinate.
Following similar calculations in the previous sections, we then obtain that
\begin{equation}
    \mathrm{Tr}{(\rho_A}^n) = \sum_{\substack{a_1,\dots ,a_n\\ b_1, \dots ,b_n}} \prod_{i=1}^n  \mathcal{O}_{0a_i} \mathcal{O}^*_{0b_i} \delta_{a_1 a_i} \delta_{a_1 b_i} \Big(\frac{D}{d_{a_1}} \Big)^{2n-2} = \sum_{a} \frac{|\mathcal{O}_{0a}|^{2n}}{(S_{0a})^{2n-2}}. 
\end{equation}
By taking the limit $n \rightarrow 1$, we obtain the TEE.
One interesting observation is that if we define $\psi_a = \mathcal{O}_{0a}$, the TEE is given by
\begin{equation}\label{twistTEE}
    S_A = \sum_a 2|\psi_a|^2 (\ln d_a - \ln |\psi_a|) -2 \ln D.
\end{equation}
Due to the unitarity of $\mathcal{O}$, we have the normalization condition $\sum_a |\psi_a|^2 = 1$.
Therefore, Eq.~(\ref{twistTEE}) is exactly the same result as discussed in Ref.~\cite{EdgeTEE} for a generic ground state in the edge theory approach. That is, although we start with a vacuum state, the twisted bipartition induces an effective state which is no longer vacuum. In general, if we start with a state $\ket{\psi} = \sum_{a} \psi_a \ket{a}$ with a torus-knot bipartition, which can be rotated back to the canonical bipartition by an operator $\mathcal{O}$, the effective state is then given by $\ket{\psi'} = \sum_{a,b} \mathcal{O}_{ab} \psi_b \ket{a}$.
For our $K(8,3)$ example where the initial state has no Wilson lines insert, the effective state is given by $\ket{\psi'} = \sum_{a,b} (ST)^{-1}_{ab} \delta_{0b} \ket{a} = \sum_{a} T^*_{aa} S_{0a} \ket{a}$. The topological entanglement entropy is then given by
\begin{equation}
    S_A = \sum_a 2(S_{0a})^2 (\ln d_a - \ln S_{0a}) -2 \ln D = 0.
\end{equation}

We have shown that the TEE for an arbitrary ground state and an arbitrary torus-knot bipartition can be written as Eq.~(\ref{decompTEE}), where the ground state TEE is defined in this case (the torus-knot bipartition) by
\begin{equation}
\label{Eq:gs}
    S_{\mathrm{gs}} := \sum_a 2|\psi_a|^2 (\ln d_a - \ln |\psi_a|),
\end{equation}
where $\ket{\psi} = \sum_{a} \psi_a \ket{a}$ depends on the twists of bipartition.
This ground state TEE is bounded by\footnote{This fact has also been mentioned in \cite{EdgeTEE} for a generic ground states. We extend their result to generic ground states in generic bipartitions. A proof can be seen in appendix~\ref{appendixB}}.
\begin{equation}\label{estimate}
    0 \leq S_{\mathrm{gs}} \leq 2\ln D.
\end{equation}
That is, in the context of TQFT, $-2 \ln D$ shown in Refs.~\cite{TEE, TEE2} is the lower bound for the TEE, independent of bipartitions, the types of anyons, and the ground states,  as long as the number of interfaces is fixed. The $S_{\mathrm{gs}}$ depends on the ground state and the (Dehn) twist of bipartition, will always be non-negative.

\section{Conclusion}
\label{Sec:5}

In this paper, we apply the invariant property of the entanglement quantities under coordinate transformations for generic torus-knots bipartitions. We derive Verlinde-like formulas using this invariant property. We find the TEEs of torus-knot bipartitions with twists can be decomposed into a state independent part which depends only on the interface and a non-negative correction caused by effective Wilson lines inserted into the system.
As a final remark, in the follow-up paper~\cite{CYL}, we will discuss a general decomposition of the TEE
\begin{equation}\label{decomp}
S_{\mathrm{TEE}}(A,\psi) = S_{\mathrm{min}}(A) + S_{\mathrm{gs}}(A,\psi),
\end{equation}
where
\begin{equation}\label{minTEE}
S_{\mathrm{min}}(A) = \min_{\ket{\psi}} S_{\mathrm{TEE}}(A,\ket{\psi}) \leq 0,
\end{equation}
is an universal topological quantity that depends only on the number of interfaces of the bipartition, and $S_{\mathrm{gs}} \geq 0$ is a quantity that detects effective ground states caused by the twisted bipartition or the Wilson lines inserted in the system. Overall, the value of this correction term will never exceed the absolute value of the universal term, ensuring that the total TEE, i.e., $S_{\mathrm{TEE}}$ is always non-positive. Furthermore, we will show that $S_{\mathrm{gs}}$ behaves just like the usual entanglement entropy, as it is non-negative and satisfies both the strong subadditivity and the subadditivity~\cite{CYL}. $S_{\mathrm{TEE}}$, on the other hand, is always non-positive and only satisfies the strong subadditivity.

\section{Acknowledgments}
We are grateful to Xueda Wen for useful discussions. P.-Y.C. acknowledges support from the National Science and Technology Council of Taiwan under Grants No. NSTC 112-2636-M-007-007 and No. 112-2112-M-007-043. Both P.-Y.C and C.-Y. L. thank the National Center for Theoretical Sciences, Physics Division for its support.

\newpage
\appendix
\section{The Verlinde-like formula}\label{appendixA}
Here, we derive Eq. (\ref{main}) by using the Verlinde formula. The Verlinde formula states that the fusion rules can be simultaneously diagonalized by the modular $S$ matrices \cite{Ver}
\begin{equation}
    \mathrm{dim} V^{a}_{bc} = (N_c)^a_b = (N_b)^a_c = \sum_d \frac{S_{bd}S_{cd}S^*_{da}}{S_{0d}}.
\end{equation}
One can generalize this to $n$ anyons by factorizing the fusion trees
\begin{equation}
     V^0_{a_1, \dots a_n} = \sum_{c_2,\dots c_{n-1}} \prod_{i=2}^n V^{c_i}_{c_{i-1} a_i},
\end{equation}
where $c_{n-1} = 0$ and $c_1 = a_1$.
Substituting the Verlinde formula into the above expression, we then obtain
\begin{equation}
    \mathrm{dim} V^0_{a_1, \dots a_n} = \sum_{\substack{c_2,\dots ,c_{n-1}\\ d_2, \dots ,d_n}} \prod_{i=2}^n \frac{S_{a_i d_i}S_{c_{i-1} d_i}S^*_{d_i c_i}}{S_{0 d_i}}.
\end{equation}
Sum over the internal $c_i$'s for $i=2, \dots ,n-1$ with $\sum_{c_i} S^*_{d_i c_i} S_{c_i d_{i+1}} = \delta_{d_i d_{i+1}}$, one has
\begin{equation}
\begin{aligned}
    \mathrm{dim} V^0_{a_1, \dots a_n} &= \sum_{d_2, \dots ,d_n} S_{c_1 d_2} \Big( \prod_{i=2}^{n-1} \frac{S_{a_i d_i} \delta_{d_i d_{i+1}} }{S_{0 d_i}} \Big) \frac{S_{a_n d_n} S^*_{c_{n-1} d_n}}{S_{0 d_n}}\\
    &= \sum_d \Big( \prod_{i=1}^n \frac{S_{a_i d}}{S_{0 d}} \Big) (S_{0d})^2.
\end{aligned}
\end{equation}
This is the Verlinde-like formula discussed in \cite{genVer}.
Now, substituting in $S_{0a} = S^*_{0a} = d_a$, $S_{00}^{-1} = D$ and $S^2 = C$, the charge conjugation operator, one arrive at
\begin{equation}
    \sum_{a_1, \dots ,a_n} \big( \prod_{i=1}^n d_{a_i} \big)  \mathrm{dim} V^0_{a_1, \dots a_n} = \sum_d \Big( \prod_{i=1}^n \frac{C_{0 d}}{S_{0 0} S_{0 d}} \Big) (S_{0d})^2 = D^{2n-2}.
\end{equation}
This is exactly equation Eq. (\ref{main}) by setting $n$ to $2n$ due to the presence of both $a_k$ and $b_k$.

\section{The bounds on ground state TEEs}\label{appendixB}
Here, we give a brief proof of Eq. (\ref{estimate}). Since we are extremizing under the constraint $\sum_a |\psi_a|^2 = 1$, we can employ the method of Lagrange multiplier by defining
\begin{equation}
    \Tilde{S}_{\mathrm{gs}} = \sum_a 2|\psi_a|^2 (\ln d_a - \ln |\psi_a|) - \lambda (\sum_a |\psi_a|^2 - 1). 
\end{equation}
Differentiating with respect to $|\psi_a|$ yields
\begin{equation}
    2(\ln d_a - \ln |\psi_a|) - 1 = 2\lambda,
\end{equation}
which normalizes to $|\psi_a| = S_{0a}$.
That is, the extremum of $S_{\mathrm{gs}}$ happens only if $|\psi_a| = S_{0a}$ or $|\psi_a|$ takes boundary values . Therefore, one then verify that $S_{\mathrm{gs}}$ takes minimum $S_{\mathrm{gs}} = 0$ when $|\psi_a| = \delta_{a 0}$ and takes maximum $S_{\mathrm{gs}} = 2 \ln D$ when $|\psi_a| = S_{0a}$ for all $a$.

\newpage
\bibliographystyle{jhep}
\bibliography{references}

\end{document}